\documentclass{article}

\usepackage[numbers]{natbib}
\usepackage{PRIMEarxiv}

\usepackage[utf8]{inputenc} 
\usepackage[T1]{fontenc}    
\usepackage{hyperref}       
\usepackage{url}            
\usepackage{booktabs}       
\usepackage{amsfonts}       
\usepackage{nicefrac}       
\usepackage{microtype}      
\usepackage{lipsum}
\usepackage{fancyhdr}       
\usepackage{graphicx}       
\graphicspath{{media/}}     

\usepackage{float}
\usepackage{tcolorbox}
\tcbuselibrary{listings}
\usepackage{tcolorbox}
\usepackage{enumitem}
\usepackage{amsmath} 
\usepackage{esint}
\usepackage{amssymb}

\def\tsc#1{\csdef{#1}{\textsc{\lowercase{#1}}\xspace}}
\tsc{WGM}
\tsc{QE}
\tsc{EP}
\tsc{PMS}
\tsc{BEC}
\tsc{DE}

\title{Electroactive differential growth and delayed instability in accelerated healing tissues
\thanks{\textit{\underline{Citation}}: 
\textbf{Yafei Wang, Zhanfeng Li, Xingmei Chen, Yun Tan, Fucheng Wang, Yangkun Du, Yunce Zhang, Yipin Su, Fan Xu, Changguo Wang, Weiqiu Chen, Ji Liu,
Electroactive differential growth and delayed instability in accelerated healing tissues,
Journal of the Mechanics and Physics of Solids,
Volume 193,
2024,
105867,
ISSN 0022-5096,
https://doi.org/10.1016/j.jmps.2024.105867.}} 
}

\author{%
  \textbf{%
    Yafei Wang$^{1,2}$, 
    Zhanfeng Li$^{3}$, 
    Xingmei Chen$^{1}$, 
    Yun Tan$^{1}$, 
    Fucheng Wang$^{1}$, 
    Yangkun Du$^{4}$\thanks{Corresponding author: \texttt{duyangkunzju@gmail.com}}
  }\\[0.8ex]
  \textbf{%
    Yunce Zhang$^{7}$, 
    Yipin Su$^{5,6}$, 
    Fan Xu$^{2}$\thanks{Corresponding author: \texttt{fanxu@fudan.edu.cn}}, 
    Changguo Wang$^{8}$\thanks{Corresponding author: \texttt{wangcg@hit.edu.cn}}
  }\\[0.8ex]
 \textbf{  Weiqiu Chen$^{5,7}$, 
  Ji Liu$^{1}$\thanks{Corresponding author: \texttt{liuj9@sustech.edu.cn}}
  } \\[1.5em]
  $^1$Department of Mechanical and Energy Engineering, Southern University of Science and Technology, Shenzhen, China\\
  $^2$Institute of Mechanics and Computational Engineering, Fudan University, Shanghai, China\\
  $^3$School of Civil Engineering and Transportation, South China University of Technology, Guangzhou, China\\
  $^4$Department of Civil, Environmental and Mechanical Engineering, University of Trento, Italy\\
  $^5$Soft Matter Research Center, Department of Engineering Mechanics, Zhejiang University, Hangzhou, China\\
  $^6$Qihang Union \& Innovation Center, Huanjiang Laboratory, Zhuji, China\\
  $^7$Center for Soft Machines and Smart Devices, Huanjiang Laboratory, Zhuji, China\\
  $^8$Center for Composite Materials and Structure, Harbin Institute of Technology, Harbin, China
}

\begin{document}
\maketitle

\begin{abstract}
Guided by experiments contrasting electrically accelerated recovery with natural healing, this study formulates a model to investigate the importance of electroactive differential growth and morphological changes in tissue repair. It underscores the clinical potential of leveraging electroactive differential growth for improved healing outcomes.
The study reveals that voltage stimulation significantly enhances the healing and growth of biological tissues, accelerating the regeneration process across various growth modalities and steering towards isotropic growth conditions that do not favor any specific growth pathways.
Enhancing the electroelastic coupling parameters improves the efficacy of bioelectric devices, initiating contraction and fortification of biological tissues in alignment with the electric field. This process facilitates swift cell migration and proliferation, as well as oriented growth of tissue. In instances of strain stiffening at elevated strains, the extreme critical growth ratio aligns with the predictions of neo-Hookean models. Conversely, for tissues experiencing strain stiffening under moderate to very low strain conditions, the strain stiffening effect substantially delays the onset of electroelastic growth instability, ultimately producing a smooth, hyperelastic surface devoid of any unstable morphologies.
Our investigation, grounded in nonlinear electroelastic field and perturbation theories, explores how electric fields influence differential growth and instability in biological tissues. We examine the interactions among dimensionless voltage, internal pressure, electroelastic coupling, radius ratio, and strain stiffening, revealing their effects on promoting growth and delaying instability. This framework offers insights into the mechanisms behind electroactive growth and its instabilities, contributing valuable knowledge to the tissue healing.
\end{abstract}




\begin{tcolorbox}[colback=white, sharp corners, boxrule=0.8pt]
	\textbf{Nomenclature}
	\begin{description}[itemsep=0.0pt]
		\item[$\mathcal{B} _r, \mathcal{B} _g, \mathcal{B} _c$] \quad Reference state, virtual grown state, and current state in the configuration evolution
		\item[$\mathbf{X}, \mathbf{x}$] \quad Coordinates of a representative material point in the reference and current states
		\item[$\mathbf{F}, \mathbf{G}, \mathbf{A}$] \quad Total deformation gradient tensor, growth tensor, and purely elastic deformation tensor
		\item[$J, J_g, J_a$] \quad Relative volume changes arising from total deformation, growth, and elastic deformation
		\item[$\mathbf{b}, \mathbf{c}$] \quad Left and right Cauchy-Green tensors associated with the total deformation gradient tensor
		\item[$\mathbf{b}_a, \mathbf{c}_a$] \quad Left and right Cauchy-Green tensors associated with the elastic deformation tensor
		\item[$\mathbf{E}, \mathbf{D}, \mathbf{P}, \varepsilon , \rho $] \quad Electric field, electric displacement, polarization, dielectric permittivity and free charge density
	\end{description}
	\begin{description}[itemsep=0.0pt]
		\item[$\mathbf{E}_l, \mathbf{E}_g$] \quad Electric field relative to the reference state and virtual grown state
		\item[$\mathbf{D}_l, \mathbf{D}_g$] \quad Electric displacement corresponding to the reference state and virtual grown state
		\item[$\mathbf{S}, \boldsymbol{\tau }, q, C$] \quad Nominal stress tensor, Cauchy stress tensor, Lagrange multiplier, and scalar function
		\item[$\tau _{rr}, \tau _{\theta \theta}, \tau _{\phi \phi}$] \quad Components of Cauchy stress in radial, circumferential, and azimuthal directions
		\item[$I_1, I_2, I_3, K_4, K_5, K_6$] \quad Six independent invariants of electroactive biomaterials
		\item[$E_R, E_r$] \quad Radial electric field in reference, and current states
		\item[$D_R, D_r$] \quad Radial electric displacement in reference, and current states
		\item[$\overline{D}_R, \overline{D}_r$] \quad Dimensionless radial electric displacement in reference and current states
		\item[$\varOmega ^*, \omega , \omega ^*$] \quad Total Helmholtz free energy density, and its elastic and electroelastic components
		\item[$\overline{\varOmega }^*, \overline{\omega }, \overline{\omega }^*$] \quad Dimensionless Helmholtz free energy density and its elastic and electroelastic components
		\item[$\lambda , \lambda _i, \lambda _o$] \quad Ratios of post- to pre-deformation radius, inner radius, and outer radius, respectively
		\item[$\alpha , \beta ; J_m$] \quad Two distinct electroelastic coupling parameters; Strain stiffening level
		\item[$\overline{V}, \overline{P}, \overline{\boldsymbol{\tau }}$] \quad Dimensionless voltage, dimensionless pressure difference, and dimensionless Cauchy stress
		\item[$\gamma =\eta \ne 1; \gamma =1,\eta \ne 1; \gamma \ne 1,\eta =1$] \quad Growth categorized by factors into isotropic, area, and fiber types
		\item[$\tilde{\chi}, \chi ^{\left( 1 \right)}, \epsilon $] \quad First-order kinematic relations, incremental motion function, and a small perturbation parameter
		\item[$\mathbf{F}^{\left( 1 \right)}, \tilde{\mathbf{A}}$] \quad Incremental displacement gradient and incremental pure elastic deformation tensor
		\item[$\mathbf{S}^{\left( 1 \right)}, \mathbf{E}_{l}^{\left( 1 \right)}$] \quad Incremental forms of the nominal stress tensor and the Lagrangian electric field
		\item[$\mathbf{S}_{0}^{\left( 1 \right)}, \mathbf{E}_{l0}^{\left( 1 \right)}$] \quad Push-forward expressions for incremental nominal stress tensor and Lagrangian electric field
		\item[$\mathcal{A} _{\alpha i\beta k}^{*}, \varGamma _{\alpha i\beta}^{*}, \mathcal{K} _{\alpha \beta}^{*}$] \quad Fourth-order, third-order, and second-order electroelastic moduli tensors
		\item[$\mathcal{A} _{0jilk}^{*}, \varGamma _{0jik}^{*}, \mathcal{K} _{0ij}^{*}$] \quad Push-forward of electroelastic moduli tensors from fourth to second orders
		\item[$\mathbf{D}_{g0}^{\left( 1 \right)}; \varPhi ^{\left( 1 \right)}$] \quad Push-forward of incremental electric displacement; incremental electric potential
		\item[$u, v, w$] \quad Incremental displacement components in radial, circumferential, and azimuthal directions
		\item[$S_{0rr}^{\left( 1 \right)}, S_{0r\theta}^{\left( 1 \right)}$] \quad Push-forward components for incremental nominal stress tensor  in $rr$ and $r\theta $ directions
		\item[$D_{g0r}^{\left( 1 \right)}$] \quad Radial component of push-forward of incremental electric displacement
		\item[$q^{\left( 1 \right)}; \mathbb{P} _m\left( x \right) $] \quad Incremental Lagrange multiplier; Legendre polynomial of order $m$
		\item[$\mathcal{U} , \mathcal{S} , \mathcal{Y} , \mathcal{G} $] \quad Displacement vector, traction vector, incremental electroelastic Stroh vector, and Stroh matrix
		\item[$\mathcal{Z} ^i, \mathcal{Z} ^o; \overline{\mathcal{Z} }^i, \overline{\mathcal{Z} }^o$] \quad Inner and outer surface impedance matrices; their respective dimensionless forms
		\item[$\overline{\mathcal{G} }; \overline{\mathcal{A} }_{0piqj}^{*}, \overline{\varGamma }_{0piq}^{*}, \overline{\mathcal{K} }_{0ij}^{*}$] \quad Dimensionless Stroh matrix; three dimensionless electroelastic moduli tensors
		\item[$\overline{\mathbb{K} }_{uv}, \overline{\mathcal{U} }$] \quad Incremental displacement ratio and dimensionless  displacement vector
		\item[$\mathring{g}_i, \mathring{\mathbb{G}}_v, \tilde{g}_{cr}^{\eta \gamma}, m_{cr}$] \quad Growth rate, growth rate ratio, critical differential growth ratio and instability mode number
		\item$\left[ \tilde{g}_{cr}^{\eta \gamma} \right] _{Ext}$; $\varepsilon _{rel}$ \quad Extreme Critical Growth Ratio (CGR); Relative dielectric permittivity
		\item[$\mathring{\mathbb{G}}_v=1; \mathring{\mathbb{G}}_v<1; \mathring{\mathbb{G}}_v>1$] \quad General isotropic growth; General area growth; General fiber growth
	\end{description}
\end{tcolorbox}

\section{Introduction}

Biological tissues display a diminishing propensity for growth, healing, and regeneration with increasing evolutionary complexity among species, attributed to the high energy demand for cellular processes and structural redundancy, leading to reduced regenerative capabilities \citep{rodrigues2019wound, gardner2022reduced}. 
This highlights the importance of advancing our understanding of the principles and materials science essential for tissue dynamics, involving complex biochemical and biomechanical interactions \citep{pena2024cellular}. 
Internal and external factors drive changes in tissue mass and morphology, impacting stress and strain through cellular mechanisms such as division and apoptosis, thereby influencing tissue morphology in response to stimuli \citep{xue2016biochemomechanical, du2020electro}. 
These developments, along with aging, diseases, and treatments, compromise tissue regenerative functions, prompting the development of bioelectric devices (see Fig.~\ref{Fig-A1}) leveraging morpho-electroelasticity to enhance healing \citep{zhang2022tough, chen2022hydrogel, wang2022wearable, wang2022mechanics, wang2021buckling, shirzaei2023stretchable}. 
Integrating discrete states and growth tensors within the deformation gradient tensor, using Kröner-Lee decomposition \citep{rodriguez1994stress}, offers a method to distinguish growth from morpho-electroelasticity deformations, addressing local incompatibilities and global integrity through differential growth \citep{taber1995biomechanics}. 
However, local incompatibilities \citep{goriely2017mathematics, lee2021geometry, liu2022geometrical} depend not only on the evolutionary form of the growth tensor but also on the electrotaxis application techniques \citep{shaner2023bioelectronic, yang2022improved, song2023bioresorbable}, differential growth nuances \citep{amar2005growth, moulton2020morphoelastic, riccobelli2020surface}, and the inherent nonlinearity of tissue materials \citep{destrade2009bending, vatankhah2017mimicking, wang2023strain}.

\begin{figure}[t!]
	\begin{center}
		\includegraphics[width=0.918\textwidth]{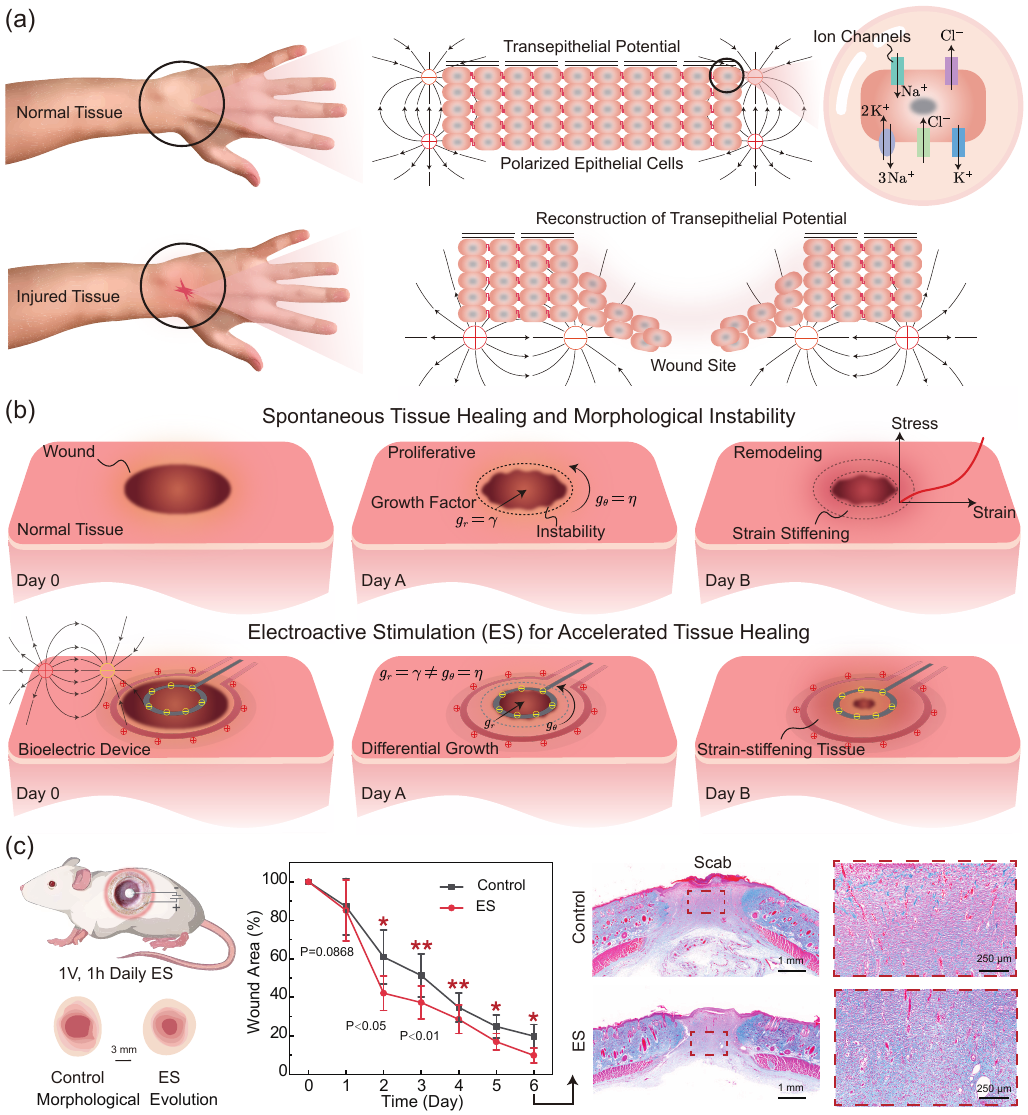}
		\caption{(a) Transepithelial Potential (TEP) and skin wound electric fields.
			First Line: TEP in intact epidermis results from ion transport, leading to a higher electrical potential at the base of the epithelium compared to the top.
			Top Right of the First Line: Asymmetric ion channels in epithelial cells.
			Second Line: Skin wound electric fields result from TEP disruption, directing potential towards the wound with a positive pole at the edge and negative at the site.
			(b) Illustration of the dichotomy between spontaneous tissue healing and its acceleration through electroactive stimulation. Note that differential growth, strain stiffening, and the morphology instability are intrinsic to this electroelastic growth process. Differential growth manifests as differences among various growth factors, while strain stiffening mainly stems from enhanced and consolidated collagen in regenerating tissues.
			(c) Electroactive stimulation accelerates tissue healing in SD rat (ES group: 1V voltage, 1 hour daily).
			Left: Schematic of wound healing in rats treated with electroactive stimulation.
			Center: Quantification of wound area for the control wounds and wounds treated with electroactive stimulation, n=4 rats in each group.
			Right: Representative images of Masson staining of full-thickness skin wounds after 6 days of electrical stimulation treatment.
		}\label{Fig-intro}
	\end{center}
\end{figure}

Electrotaxis, the guidance of cells by electric fields, showcases the complex interaction between electric fields and cellular behavior through electroelastic coupling, affecting cell deformation and migration across various cell types like monocytes, macrophages, endothelial cells, and fibroblasts \citep{lin2008lymphocyte, sun2019infection, ammann2021vascular, brown1994electric}. This behavior stems from the asymmetric distribution of ion channels in cell membranes (see Fig.~\ref{Fig-intro}a, top right of first line), creating transepithelial potentials essential for tissue healing \citep{luo2021accelerated, zhao2009electrical}. Wounds can induce the reorganization of the surrounding electric fields, which guide cell migration and influence morphology, as depicted in Fig.~\ref{Fig-intro}a.
Endogenous electric field efficacy in healing varies with individual factors such as age and health, whereas external electrical stimulation has been employed to enhance tissue repair, mimicking natural electric field effects to accelerate regeneration, as detailed in Fig.~\ref{Fig-intro}b, promoting processes like angiogenesis and cellular proliferation \citep{nuccitelli2011electric, garcia2019energy, sebastian2011acceleration, de2015immediate, mao2020biodegradable}. 
This burgeoning field of research highlights the potential of electrotaxis and electroelastic coupling in tissue healing, advocating a comprehensive approach to augment repair mechanisms. The interaction between electrotaxis and tissue growth, particularly how electric field strength and distribution affect morpho-electroelastic biomechanics, represents a valuable yet underexplored avenue for enhancing healing strategies through electric field modulation.

More importantly, under the influence of electric fields, the complex morphogenesis of biological tissues often occurs through differential growth effects \citep{du2018modified}, induced by the inherent heterogeneity of the growth field.
This process is referred to as electroactive differential growth \citep{li2023numerical}. 
Electric fields direct epithelial migration toward wound sites, as detailed in Fig.~\ref{Fig-intro}b, disrupting the initial configurations of wounds and orchestrating cellular behavior and the healing process \citep{lei2021conductive, wang2022endogenous}.
Observations of wound healing (see Fig.~\ref{Fig-intro}b), particularly the undulation of wound borders in vivo, reveal the interplay between electroactive differential growth ($g_r \ne g_{\theta} $), tissue stress, and tissue non-linearity properties, challenging traditional healing models \citep{hamed2011fibronectin, ben2013anisotropic}. The proliferation of keratinocytes during re-epithelialization in large and irregularly shaped wounds exemplifies the critical role of electroactive differential growth in navigating the complex tissue environment, influenced by residual stresses and the surrounding mature tissue framework \citep{bayly2014mechanical, theocharidis2022strain}. These dynamics affirm the impact of electroactive differential growth on regeneration, suggesting a need for precise electric field modulation during scar remodeling to ensure effective tissue healing.
The domain of differential growth modeling is categorized into two main segments: the first involves growth driven by anisotropic elastic properties, where the configuration of growth tensor is determined by the compliance tensor, favoring expansion in less stiff directions \citep{braeu2019anisotropic, soleimani2020novel}. 
The second addresses microstructures like longitudinal fibers influencing growth direction through a growth tensor aligned with their orientation, elucidating processes such as brain cortical folding \citep{lubarda2002mechanics, menzel2005modelling, bayly2014mechanical}.
Notably, the realm of electroactive differential growth remains largely uncharted, underlining a critical need for developing mechanics theories in this area, thus opening new pathways for research.

Beyond electroactive differential growth, soft tissue often exhibits a typical nonlinear strain-stiffening effect, which biologically stems from the directional contraction, rotation, and alignment of tissue cells, forming the basis of their self-protection and tissue growth and remodeling \citep{weiss1989exogenous, ambic1993influence, thawer2001effects, danjo1998actin,harmansa2023growth, kumar2024balancing, vatankhah2017mimicking, rodrigo2021engineered, wang2023strain}. Experimental research has underscored the significant contribution of semiflexible polymer networks to strain-stiffening biomimetic gels for tissue growth \citep{storm2005nonlinear} and emphasized the fundamental role of strain-controlled criticality in active fiber networks for tissue remodeling \citep{sharma2016strain}. Moreover, studies on post-mortem brain tissue stiffening reveal the effects of dehydration and microstructural changes, stressing the importance of understanding the complex mechanics of tissue nonlinearity and the impact of microstructure on tissue growth and remodeling processes \citep{weickenmeier2018brain}.
However, it is worth noting that, in addition to adjusting the endogenous composition of the material, exogenous electrical stimulation can significantly impact its strain-stiffening properties. For accelerated tissue healing, electrical stimulation has been shown to effectively modulate strain stiffening, optimizing myofibroblast contraction during re-epithelialization to minimize hypertrophic scars in mice and enhancing keratinocyte activity in adult wound healing, thereby expediting tissue closure and reducing fibrosis potential \citep{yang2022improved, song2023bioresorbable, hunckler2017current}. Targeting myofibroblast apoptosis post-re-epithelialization through electroactive interventions promises to mitigate fibrocontractile disorders, underscoring the potential to adjust the healing environment and curtail excessive scar tissue formation \citep{gabbiani2003myofibroblast}. Furthermore, employing electroactive stimulation in extensive wounds, such as burns, offers a strategic approach to counteract the adverse effects of strain stiffening by tailoring healing processes to diverse mechanical stresses, thus reducing scar tissue and fibrosis \citep{liang2020application, so2020synergistic}. Therefore, this method, complementing endogenous bioelectric fields, represents a significant advancement in healing strategies by controlling strain stiffening to foster more efficient and less fibrotic healing outcomes.

Despite significant efforts, the development of exogenous electrical stimulation devices tailored for biological tissues remains at the forefront of research, with a particular focus on unraveling electroactive differential growth and electroelastic coupling mechanisms. 
Central to this field is the refinement of electrotaxis for controlled differential growth, highlighting the necessity for a thorough exploration of morpho-electroelastic mechanisms. 
Despite having theoretical foundations in nonlinear electroelasticity and volume growth, the comprehensive effects of electric fields on tissue healing, differential growth, and strain stiffening, especially regarding their stability and interaction dynamics, remain largely unexplored. 
Addressing these gaps is crucial for understanding the intricate mechanics underlying tissue regeneration and for the development of effective healing strategies.

As shown in Fig.~\ref{Fig-intro}c and Fig.~\ref{Fig-A2}, 
this study is inspired by observations of spontaneous healing (Control) in SD rat following 8-mm skin punches (BIOPSY PUNCH) and the augmented tissue regeneration (ES) achieved through external electrical stimulation with 3D-printed conductive polymer (Ag ink, BASE-SCD2, Prtronic) . All animal surgeries are approved by the Committee on Animal Care at SUSTech, Protocol No. SUSTech-JY202312002.
In the experiment, the wound morphology of both the ES and Control groups undergoes various degrees of evolution (see Fig.~\ref{Fig-intro}c, left). Notably, the wound area ($\%$) of the rats in the ES group, which receives daily electrical stimulation at 1V for one hour (Note that the electrochemical stability window of our portable battery is 0 to 2.5V.), is significantly smaller than that observed in the Control group (see Fig.~\ref{Fig-intro}c, center).
Wherein, the statistical significance with a $\mathrm{P}$-value is less than 0.05 after the first day (For P-values, P<0.05 indicates significance (*), and P<0.01 indicates extreme significance (**)).
Masson-stained tissue sections (see Fig.~\ref{Fig-intro}c, right) reveal that the tissue slice width in the ES group is significantly reduced compared to the Control group. In the ES group, the healing tissue displays a higher density of collagen fibers and lacks notable scab formation.
To uncover the mechanisms behind these experimental observations and elucidate the role of electroactive differential growth in accelerated tissue healing, this study examines the synergy between area and fiber growth, electroelastic coupling, and the influences of voltage, pressure, and strain stiffening on tissue repair. 
Due to restrictions related to animal ethics, creating deep and large wounds significantly increases the pain levels in SD rats. Therefore, in our experiment, we only removed the full-thickness skin layer of the SD rats. To model a complex scenario, such as infected chronic wounds, we represent the 3D wound as an incompressible, hyperelastic spherical shell subjected to variable thickness and distinct voltage and pressure gradients \citep{valero2015modeling, dorfmann2014nonlinear}. This approach allows us to successfully simulate the dynamic growth and regeneration process under exogenous electrical stimulation.
Moreover, using perturbation analysis, we showcase how nonlinear strain stiffening contributes to growth instability, enhancing the understanding of electroactively accelerated tissue growth.

This paper is organized as follows: Section \ref{section2} explores electroactive tissue growth, applying the Kröner-Lee decomposition to the deformation gradient tensor and detailing the transformation relationships between electric fields and displacements within Lagrangian and Eulerian descriptions. It includes the derivation of the Cauchy stress tensor with electroelastic coupling. 
Section \ref{section3} reveals the base state of electroelastic differential growth and simplifies the first Cauchy equation, leading to the decoupling of Helmholtz free energy density and formulas for dimensionless voltage and electric displacement.
In Section \ref{section4}, we discuss generalized incremental equilibrium equations and incremental Maxwell's equations, crucial for analyzing symmetry breaking and bifurcation in electroelastic growth, utilizing Stroh formulation. Section \ref{section5} examines numerical results, focusing on the dynamics of electroelasticity-driven differential growth, its impact on accelerated healing, and delayed instability due to strain stiffening.
Finally, Section \ref{section6} summarizes our findings, highlighting their significance for enhancing tissue healing and laying groundwork for future work in biotissue engineering and regenerative medicine.

\begin{figure}[!t]
	\begin{center}
		\includegraphics[width=0.92\textwidth]{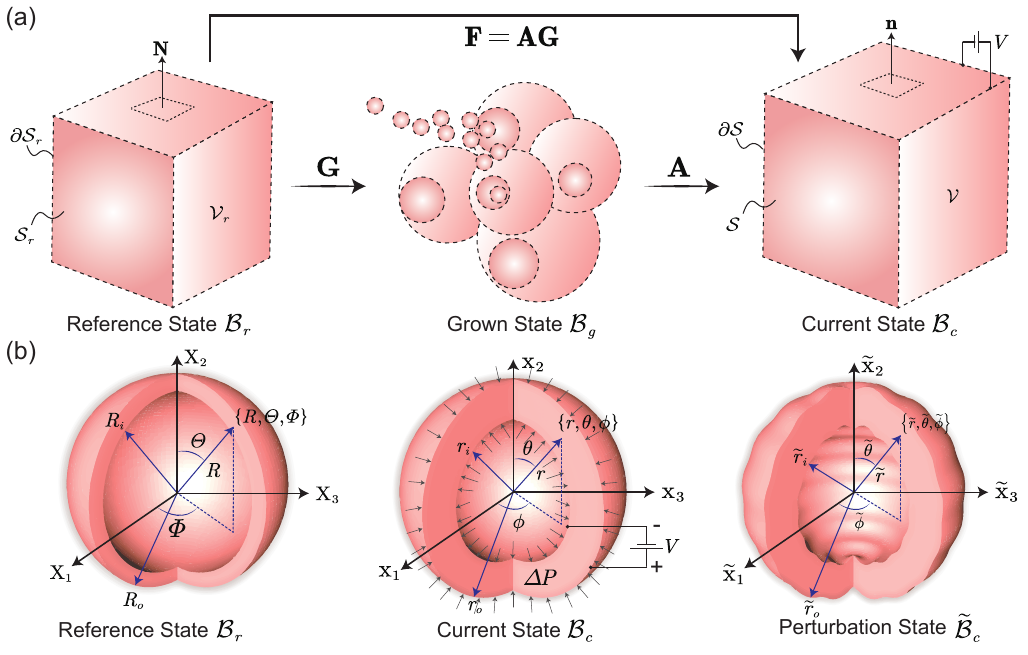}
		\caption{
			(a) Configuration evolution of electroactive differential growth materials and the multiplicative decomposition diagram of the overall deformation gradient tensor.
			(b) Geometric characterization of the healing configuration in biotissues exhibiting electroactivity and electroelasticity, with pressure applied to internal and external surfaces.
		}\label{Fig-growth}
	\end{center}
\end{figure}

\section{Electroelastic growth model}\label{section2}
When examining the growth of a solid material that exhibits finite electroelasticity deformation \citep{dorfmann2014nonlinear} under external biasing fields, we note a transformation: the position vector, originally in the reference state $\mathcal{B} _r$ as $\mathbf{X}$, evolves into $\mathbf{x}$ within the current state $\mathcal{B} _c$. 
This transformation is articulated via the bijective motion mapping given by $\mathbf{x}=\chi (\mathbf{X},t)$. Consequently, the deformation gradient tensor can be expressed through the gradient of this motion, represented as $\mathbf{F}=\mathrm{Grad}\chi ={{\partial \mathbf{x}}/{\partial \mathbf{X}}}$. 
Additionally, the Jacobian, symbolized as $J=\det \mathbf{F}$, remains consistently positive throughout all deformations. 
To encapsulate the intertwined effects of inherent stress during growth \citep{amar2005growth, goriely2017mathematics, du2020electro, wang2023strain}, we employ a Kröner-Lee decomposition delineated as $\mathbf{F}=\mathbf{AG}$ (see Fig. \ref{Fig-growth}a). 
Here, the growth tensor $\mathbf{G}$ shifts $\mathcal{B} _r$ into a conceptual growth state $\mathcal{B} _g$, and the pure elastic deformation tensor $\mathbf{A}$ subsequently crafts the current configuration $\mathcal{B} _c$ from segments within $\mathcal{B} _g$. 
The Jacobian transformations corresponding to the various deformation gradients are defined as follows: $J_g=\det \mathbf{G}$ and $J_a=\det \mathbf{A}$. 
In the case of incompressible materials, we can observe that $J_a=\det \mathbf{A}=1$, and $J=\det \mathbf{F}=J_g$. 
The left and right Cauchy-Green tensors associated with $\mathbf{F}$ are denoted here by $\mathbf{b}=\mathbf{FF}^{\mathrm{T}}$ and $\mathbf{c}=\mathbf{F}^{\mathrm{T}}\mathbf{F}$, respectively, where $^{\mathrm{T}}$ represents the transpose of a tensor.

In the designated current state, $\mathcal{B} _c$, we define three vectors related to the electric field (see Fig. \ref{Fig-growth}a). The electric field itself is given by $\mathbf{E}$, its associated electric displacement is $\mathbf{D}$, and we have the polarization vector, $\mathbf{P}$. 
Among these, $\mathbf{D}$ and $\mathbf{E}$ are crucial field variables.  
When necessary, the polarization vector can be defined using the standard relationship: $\mathbf{P}=\mathbf{D}-\varepsilon \mathbf{E}$.
When we step outside this material into a vacuum, the relationship is distilled to its simplest form: $\mathbf{D}=\varepsilon \mathbf{E}$. 
Here, $\varepsilon$ denotes the dielectric permittivity.
To distinguish in different scenarios, we label the Lagrangian versions of $\mathbf{D}$ and $\mathbf{E}$ as $\mathbf{D}_l$ and $\mathbf{E}_l$, respectively. In the virtual state $\mathcal{B} _g$, these become $\mathbf{D}_g$ and $\mathbf{E}_g$.
Considering the quasi-static deformation scenario, and given the absence of magnetic fields and time dependence, the Maxwell's Eulerian field equations can be distilled to:
\begin{align}\label{Eq1}
	\begin{split}
	\mathrm{curl}\mathbf{E} = \mathbf{0}, \quad \text{and} \quad \mathrm{div}\mathbf{D} = \rho.
	\end{split}
\end{align}
where $\rho$ denotes the free charge density, $\mathrm{curl}$ and $\mathrm{div}$ stand for the curl and divergence operators in the current state, respectively. 
Importantly, these equations hold true both inside and outside of electro-sensitive materials, such as biotissue materials, provided that there are no free charges present within the material.
Parallelly, in the Lagrangian description, the Maxwell's equations can be expressed as $\mathrm{Curl}\mathbf{E}_l=\mathbf{0}$ and $\mathrm{Div}\mathbf{D}_l=\rho_l=\rho J$.
Within these representations, the operators $\mathrm{Curl}$ and $\mathrm{Div}$ correspond to the curl and divergence, respectively, relative to the reference state denoted as $\mathcal{B} _r$.

Subsequently, to elucidate the relationships among the electric fields \( \mathbf{E}_l \), \( \mathbf{E}_g \), and \( \mathbf{E} \) corresponding to the initial reference state \( \mathcal{B}_r \), the virtual growth state \( \mathcal{B}_g \), and the current state \( \mathcal{B}_c \), we turn to Eq.~\eqref{Eq1}$_1$ and integrate it across an arbitrary open surface $\mathcal{S} $ (note that $\partial \mathcal{S} $ represents the boundary curve of $\mathcal{S} $, see Fig. \ref{Fig-growth}a). Then we apply Stokes' theorem to obtain
\begin{align}\label{Eq2}
	\begin{split}
		\iint_{\mathcal{S}}{\left( \mathrm{curl}\mathbf{E} \right) \cdot \mathbf{n}}\mathrm{d}s=\oint_{\partial \mathcal{S}}{\mathbf{E}\cdot \mathrm{d}\mathbf{x}}=0.
	\end{split}
\end{align}
Considering the line element \( \mathrm{d}\mathbf{x} \), it originates from the corresponding line element \( \mathrm{d}\mathbf{X} \) in the reference state. 
This relationship can be expressed as $\mathrm{d}\mathbf{x} = \mathbf{F}\mathrm{d}\mathbf{X}$. 
Given this, we can transform the line integral over the deformed state to an integral over the reference boundary $\partial \mathcal{S}_r$ of the original reference surface $\mathcal{S}_r$. 
By leveraging the relation $\mathbf{E} \cdot \left( \mathbf{F}\mathrm{d}\mathbf{X} \right) = \left( \mathbf{F}^{\mathrm{T}}\mathbf{E} \right) \cdot \mathrm{d}\mathbf{X}$, 
we can then apply Stokes' theorem over the reference domain to proceed and obtain
\begin{align}\label{Eq3}
	\begin{split}
		\oint_{\partial \mathcal{S} _r}{\left( \mathbf{F}^{\mathrm{T}}\mathbf{E} \right) \cdot \mathrm{d}\mathbf{X}}=\iint_{\mathcal{S} _r}{\left( \mathrm{Curl}\left( \mathbf{F}^{\mathrm{T}}\mathbf{E} \right) \right) \cdot \mathbf{N}\mathrm{d}S}=0,
	\end{split}
\end{align}
Since the surface $\mathcal{S}_r$ is arbitrary, we have $\mathrm{Curl}\left( \mathbf{F}^{\mathrm{T}}\mathbf{E} \right) =0$. 
This indicates that, through the corresponding deformation gradient tensor $\mathbf{F}^{\mathrm{T}}$, the electric field vector $\mathbf{E}$ relative to the current state can be subjected to a pull-back operation \citep{dorfmann2014nonlinear}, transforming it into the electric field in the Lagrangian form. 
Adopting the same pull-back concept, we can derive the following relationships:
\begin{align}\label{Eq4}
	\begin{split}
	\mathbf{E}_l=\mathbf{F}^{\mathrm{T}}\mathbf{E},\quad \mathrm{and}\quad \mathbf{E}_g=\mathbf{A}^{\mathrm{T}}\mathbf{E}.
	\end{split}
\end{align}

To derive the relationships between the electric displacements \( \mathbf{D}_l \), \( \mathbf{D}_g \), and \( \mathbf{D} \) corresponding to three distinct states, we begin by integrating Eq.~\eqref{Eq1}$_2$ over an arbitrary volume $\mathcal{V} $ (where \( \partial \mathcal{V} \) denotes the regular boundary of $\mathcal{V} $). Subsequently, we utilize Nanson's formula in conjunction with the Gauss's divergence theorem to arrive at:
\begin{align}\label{Eq5}
	\begin{split}
		\iiint_{\mathcal{V}}{\mathrm{div}\mathbf{D}\mathrm{d}v}=\oiint_{\partial \mathcal{V}}{\mathbf{D}\cdot \mathbf{n}}\mathrm{d}s=\oiint_{\partial \mathcal{V} _r}{\left( J\mathbf{F}^{-1}\mathbf{D} \right) \cdot \mathbf{N}}\mathrm{d}S=\iiint_{\mathcal{V} _r}{\mathrm{Div}\left( J\mathbf{F}^{-1}\mathbf{D} \right)}\mathrm{d}V=\iiint_{\mathcal{V} _r}{\rho J \mathrm{d}V},
	\end{split}
\end{align}
Given that the volume $\mathcal{V}_r$ in reference state is arbitrary (see Fig. \ref{Fig-growth}a), it follows that $\mathrm{Div}\left( J\mathbf{F}^{-1}\mathbf{D} \right) = \rho J$.
This indicates that, by utilizing the inverse of the relevant deformation gradient tensor, we can also pull-back the electric displacement vector from the current state to its reference state.
Thus, concerning the electric displacements in the three different states, we can establish the following pivotal relationships:
\begin{align}\label{Eq6}
	\begin{split}
		\mathbf{D}_l=J\mathbf{F}^{-1}\mathbf{D},\quad \mathrm{and}\quad \mathbf{D}_g=J_a\mathbf{A}^{-1}\mathbf{D}.
	\end{split}
\end{align}

If the material is subjected to incompressibility constraints, involving one or more scalar relationships (denoted here as $C(\mathbf{A})=\det \left( \mathbf{A} \right) -1$) among the elastic strains, it becomes essential to introduce a Lagrange multiplier \(q\) during the derivation of the nominal stress tensor \(\mathbf{S}\). To accommodate these constraints, we can modify the energy function \(\tilde{\varOmega}(\mathbf{A},\mathbf{D}_g)\) by including the term \(qC(\mathbf{A})\). 
Subsequently, by leveraging the relationship \(\mathbf{A} = \mathbf{FG}^{-1}\), the total nominal stress tensor is obtained through the following relation:
\begin{align}\label{Eq7}
	\begin{split}
		\mathbf{S}=J\frac{\partial \tilde{\varOmega}\left( \mathbf{A},\mathbf{D}_g \right)}{\partial \mathbf{F}}-qJ\frac{\partial C\left( \mathbf{A} \right)}{\partial \mathbf{F}}=J\mathbf{G}^{-1}\frac{\partial \tilde{\varOmega}\left( \mathbf{A},\mathbf{D}_g \right)}{\partial \mathbf{A}}-qJ\mathbf{G}^{-1}\mathbf{A}^{-1}.
	\end{split}
\end{align}

Utilizing the relationship $\mathbf{S}=J\mathbf{F}^{-1}\boldsymbol{\tau }$, we obtain the Cauchy stress tensor $\boldsymbol{\tau }$ as follows:
\begin{align}\label{Eq8}
	\begin{split}
		\boldsymbol{\tau }=\mathbf{A}\frac{\partial \tilde{\varOmega}\left( \mathbf{A},\mathbf{D}_g \right)}{\partial \mathbf{A}}-q\mathbf{I}.
	\end{split}
\end{align}

Given that the electric field vector in the virtual growth state $\mathcal{B}_g$ is calculated as $\mathbf{E}_g = {{\partial \tilde{\varOmega}}/{\partial \mathbf{D}_g}}$, Eq.~\eqref{Eq4} facilitates obtaining the electric fields with respect to both the reference state and the current state, achieved through the following relationships:
\begin{align}\label{Eq9}
	\begin{split}
		\mathbf{E}_l=\mathbf{G}^{\mathrm{T}}\frac{\partial \tilde{\varOmega}\left( \mathbf{A},\mathbf{D}_g \right)}{\partial \mathbf{D}_g},\quad \mathrm{and}\quad \mathbf{E}=\mathbf{A}^{-\mathrm{T}}\frac{\partial \tilde{\varOmega}\left( \mathbf{A},\mathbf{D}_g \right)}{\partial \mathbf{D}_g}.
	\end{split}
\end{align}

For the sake of simplicity, the focus of this analysis is on isotropic electro-sensitive biotissue materials.
In the context of isotropic elasticity, the energy is typically a function of three independent invariants of the Cauchy-Green tensor, with the principal invariants $I_1$, $I_2$, and $I_3$ being commonly used. Conversely, in the case of isotropic electroelastic materials, the energy function is treated as an isotropic function of the corresponding right Cauchy-Green deformation tensor $\mathbf{c}_a=\mathbf{A}^{\mathrm{T}}\mathbf{A}$ (equally,  $\mathbf{b}_a=\mathbf{A}\mathbf{A}^{\mathrm{T}}$) and the outer product of the electric displacement vector $\mathbf{D}_g\otimes \mathbf{D}_g$. It depends on these quantities through six invariants, which include $I_1$, $I_2$, $I_3$, and three additional invariants that are functions of $\mathbf{D}_g$. To streamline our analysis, we adopt a reduced form of the energy function \citep{mehnert2021complete, liu2021coupled}, denoted as $\tilde{\varOmega}=\tilde{\varOmega}\left( I_1,I_2,I_3,K_4,K_5,K_6 \right)$. Here, we employ the following formulation to define the invariants:
\begin{align}\label{Eq10}
	\begin{split}
	I_1=\mathrm{tr}\mathbf{c}_a,\,\, I_2=\frac{\left( \mathrm{tr}\mathbf{c}_a \right) ^2-\mathrm{tr}\mathbf{c}_{a}^{2}}{2},\,\, I_3=\det \mathbf{c}_a,\,\, K_4=\mathbf{D}_g\cdot \mathbf{D}_g,\,\, K_5=\left( \mathbf{c}_a\mathbf{D}_g \right) \cdot \mathbf{D}_g,\,\, K_6=\left( \mathbf{c}_{a}^{2}\mathbf{D}_g \right) \cdot \mathbf{D}_g.
	\end{split}
\end{align}

In the notation below, the subscript $i$ on $\tilde{\varOmega}$ represents differentiation with respect to the $i^{th}$ invariant, where $i = 1, 2, \ldots, 6$. We aim to determine the first-order derivatives of these invariants with respect to both $\mathbf{A}$ and $\mathbf{D}_g$. Notably, the invariants $I_1$, $I_2$, and $I_3$ are independent of $\mathbf{D}_g$. As a result, we have:
\begin{align}\label{Eq11}
	\begin{split}
		&\frac{\partial I_1}{\partial \mathbf{A}}=2\mathbf{A}^{\mathrm{T}},\quad \frac{\partial I_2}{\partial \mathbf{A}}=2\left( I_1\mathbf{A}^{\mathrm{T}}-\mathbf{A}^{\mathrm{T}}\mathbf{b}_a \right) ,\quad \frac{\partial I_3}{\partial \mathbf{A}}=2I_3\mathbf{A}^{-1},\quad\frac{\partial K_5}{\partial \mathbf{A}}=2\mathbf{D}_g\otimes \mathbf{AD}_g,\\ &\frac{\partial K_6}{\partial \mathbf{A}}=2\left( \mathbf{D}_g\otimes \mathbf{Ac}_a\mathbf{D}_g+\mathbf{c}_a\mathbf{D}_g\otimes \mathbf{AD}_g \right) ,\quad \frac{\partial K_4}{\partial \mathbf{D}_g}=2\mathbf{D}_g,\quad \frac{\partial K_5}{\partial \mathbf{D}_g}=2\mathbf{c}_a\mathbf{D}_g,\quad \frac{\partial K_6}{\partial \mathbf{D}_g}=2\mathbf{c}_{a}^{2}\mathbf{D}_g.
	\end{split}
\end{align}

By meticulously examining Eq.~\eqref{Eq8} and Eq.~\eqref{Eq11}, the Cauchy stress tensor can be obtained via the chain rule, i.e.,
\begin{align}\label{Eq12}
	\begin{split}
		\boldsymbol{\tau }&=\mathbf{A}\left( \tilde{\varOmega}_1\frac{\partial I_1}{\partial \mathbf{A}}+\tilde{\varOmega}_2\frac{\partial I_2}{\partial \mathbf{A}}+\tilde{\varOmega}_5\frac{\partial K_5}{\partial \mathbf{A}}+\tilde{\varOmega}_6\frac{\partial K_6}{\partial \mathbf{A}} \right) -q\mathbf{I}\\&=2\tilde{\varOmega}_1\mathbf{b}_a+2\tilde{\varOmega}_2\left( I_1\mathbf{b}_a-\mathbf{b}_{a}^{2} \right) +2\tilde{\varOmega}_5\mathbf{D}\otimes \mathbf{D}+2\tilde{\varOmega}_6\left( \mathbf{D}\otimes \mathbf{b}_a\mathbf{D}+\mathbf{b}_a\mathbf{D}\otimes \mathbf{D} \right) -q\mathbf{I}.
	\end{split}
\end{align}

Based on Eqs.~\eqref{Eq9}$_2$ and \eqref{Eq11}, the electric field is represented by the following relation:
\begin{align}\label{Eq13}
	\begin{split}
		\mathbf{E}=\mathbf{A}^{-\mathrm{T}}\left( \tilde{\varOmega}_4\frac{\partial K_4}{\partial \mathbf{D}_g}+\tilde{\varOmega}_5\frac{\partial K_5}{\partial \mathbf{D}_g}+\tilde{\varOmega}_6\frac{\partial K_6}{\partial \mathbf{D}_g} \right) =2\left( \tilde{\varOmega}_4\mathbf{b}_{a}^{-1}+\tilde{\varOmega}_5\mathbf{I}+\tilde{\varOmega}_6\mathbf{b}_a \right) \mathbf{D}.
	\end{split}
\end{align}

Building upon the aforementioned discussions, in situations devoid of mechanical body forces and distributed free charges, the fundamental governing equations for addressing any boundary-value problem are encapsulated in the equations that follow:
\begin{align}\label{Eq14}
	\begin{split}
		\mathrm{div}\boldsymbol{\tau }=\mathbf{0}, \quad \mathrm{curl}\mathbf{E}=\mathbf{0},\quad \mathrm{and}\quad \mathrm{div}\mathbf{D}=0.
	\end{split}
\end{align}

The pertinent boundary conditions corresponding to the total Cauchy stress, electric field, and electric displacement are expressed as:
\begin{align}\label{Eq15}
	\begin{split}
		\boldsymbol{\tau  n}=\boldsymbol{\tau }_a+\boldsymbol{\tau }_{e}^{*}, \quad \mathbf{n}\times \left( \mathbf{E}-\mathbf{E}^* \right) =\mathbf{0},\quad \mathrm{and}\quad \mathbf{n}\cdot \left( \mathbf{D}-\mathbf{D}^* \right) =0.
	\end{split}
\end{align}
Here, $\boldsymbol{\tau }_a$ denotes the applied mechanical load per unit of the deformed area. 
Meanwhile, $\boldsymbol{\tau }_{e}^{*}$ symbolizes a force per unit area in the deformed configuration due to the Maxwell stress located beyond the boundary of the material \citep{dorfmann2014nonlinear}. 
The vectors $\mathbf{E}^*$ and $\mathbf{D}^*$ correspond to the external electric field and electric displacement, respectively, both assessed at the boundary. 
Notably, the surface free charge density is not considered.
Disregarding external electric fields, and considering direct contact between the bioelectric device and the wound tissue with deterministic voltage applied directly at the boundary, the terms $\boldsymbol{\tau }_{e}^{*}$, $\mathbf{E}^*$, and $\mathbf{D}^*$ in the aforementioned equation vanish.

\section{Electroelastic differential growth for strain-stiffening healing tissues}\label{section3}
\subsection{Nonlinear electroelastic growth and deformation}
Consider a thick-walled spherical shell in the reference state (see Fig. \ref{Fig-growth}b for details), denoted as $\mathcal{B}_r$, described using spherical polar coordinates $\left\{ R,\Theta ,\Phi \right\} $ : $\left\{ 0<R_i\leqslant R\leqslant R_o,0 \leqslant \Theta \leqslant \pi ,0\leqslant \Phi \leqslant 2\pi \right\} $, where $R_i$ and $R_o$ represent the inner and outer radii of the spherical shell, respectively.
In its current state, denoted as $\mathcal{B}_c$, it occupies the region: $\left\{ 0<r_i\leqslant r\leqslant r_o,0 \leqslant \theta \leqslant \pi ,0\leqslant \phi \leqslant 2\pi \right\} $.

Voltage is applied to both the inner and outer surfaces of the shell, with a pressure difference $\varDelta P$ existing between these surfaces.
To simplify the problem, the spherical shell undergoes radial and symmetrical inflation (or deflation) by applying a pressure \(P\) on its inner surface, resulting in a spherically symmetric deformation. 

Subsequently, the deformation gradient tensor, when referred to the two sets of spherical polar coordinate axes, is represented by the matrix \(\mathbf{F}\), expressed as $\mathbf{F}=\mathrm{diag}\left( {{\partial r}/{\partial R}},\lambda ,\lambda \right) $, where $\lambda ={{r}/{R}}$. Next, we postulate that this geometric deformation results from a combination of growth deformation \(\mathbf{G}\) and elastic deformation \(\mathbf{A}\) (as discussed in the previously mentioned Kröner--Lee decomposition, see Fig. \ref{Fig-growth}). Furthermore, we assume that both the growth tensor and the pure elastic deformation tensor maintain spherical symmetry, allowing us to represent the tensors \(\mathbf{G}\) and \(\mathbf{A}\) as follows:
\begin{align}\label{Eq16}
	\begin{split}
		\mathbf{G} = \begin{bmatrix}
			\gamma & 0 & 0 \\
			0 & \eta & 0 \\
			0 & 0 & \eta
		\end{bmatrix},
		\quad \mathrm{and}\quad
		\mathbf{A} = \begin{bmatrix}
			\gamma^{-1}\frac{\partial r}{\partial R} & 0 & 0 \\
			0 & \eta^{-1}\lambda & 0 \\
			0 & 0 & \eta^{-1}\lambda
		\end{bmatrix}.
	\end{split}
\end{align}
where \(\gamma = J_g \eta^{-2}\), \(\gamma\) is the radial growth factor, and both the circumferential and azimuthal growth factors are equal to \(\eta\).
Following this, when the incompressibility condition is imposed on tensor $\mathbf{A}$, the deformation is exclusively determined by ${{\partial r}/{\partial R}}=J_g\lambda ^{-2}$. Integrating this equation yields the following significant geometric relationship:
\begin{align}\label{Eq17}
	\begin{split}
		\det \mathbf{A}=1 \quad \Longrightarrow  \quad \,r^3=r_{i}^{3}+3\int_{R_i}^R{J_g\left( R \right) R^2\mathrm{d}R},
	\end{split}
\end{align}
where $r=r\left( R \right) $. If the growth factors do not evolve with deformation, the relationship $r^3=r_{i}^{3}+J_g\left( R^3-R_{i}^{3} \right)$ holds. Therefore, the principal invariants with respect to \(I_1\) and \(I_2\) can be expressed as:
\begin{align}\label{Eq18}
	\begin{split}
		I_1=J_{g}^{2}\gamma ^{-2}\lambda ^{-4}+2\eta ^{-2}\lambda ^2,\quad \mathrm{and}\quad I_2=2J_g\gamma ^{-1}\!\:\lambda ^{-2}+\lambda ^4\eta ^{-4}.
	\end{split}
\end{align}

Given the spherical symmetry, we restrict our analysis to radial electric fields and electric displacements. Building upon Eq. \eqref{Eq4}, we can express the radial electric field and electric displacement in the context of the reference state $\mathcal{B}_r$, virtual growth state $\mathcal{B}_g$, and current state $\mathcal{B}_c$ through the following formalisms:
\begin{align}\label{Eq19}
	\begin{split}
	\mathbf{E}_l=\left[ E_R,0,0 \right] ^{\mathrm{T}},\,\,\mathbf{E}_g=\mathbf{G}^{-\mathrm{T}}\mathbf{E}_l=\left[ \gamma ^{-1}E_R,0,0 \right] ^{\mathrm{T}},\,\,\mathbf{E}=\left[ E_r,0,0 \right] ^{\mathrm{T}}=\left[ J_{g}^{-1}\lambda ^2E_R,0,0 \right] ^{\mathrm{T}},\\\mathbf{D}_l=\left[ D_R,0,0 \right] ^{\mathrm{T}},\,\,\mathbf{D}_g=J_{g}^{-1}\mathbf{GD}_l=\left[ \eta ^{-2}D_R,0,0 \right] ^{\mathrm{T}},\,\,\mathbf{D}=\left[ D_r,0,0 \right] ^{\mathrm{T}}=\left[ \lambda ^{-2}D_R,0,0 \right] ^{\mathrm{T}}.
	\end{split}
\end{align}
from which the invariants corresponding to $K_4$, $K_5$, and $K_6$ can be derived using the subsequent relations:
\begin{align}\label{Eq20}
	\begin{split}
		K_4=D_{R}^{2}\eta ^{-4}, \quad K_5=D_{R}^{2}\lambda ^{-4}, \quad K_6=D_{R}^{2}\eta ^4\lambda ^{-8}.
	\end{split}
\end{align}

Utilizing Eq. \eqref{Eq12}, the components of the Cauchy stress tensor — specifically, $\tau _{rr}$, $\tau _{\theta \theta}$, and $\tau _{\phi \phi}$ — can be determined as:
\begin{align}\label{Eq21}
	\begin{split}
		&\tau _{rr}=2\!\:\eta ^4\lambda ^{-4}\tilde{\varOmega}_1+4\!\:\eta ^2\lambda ^{-2}\tilde{\varOmega}_2+2\lambda ^{-4}\tilde{\varOmega}_5D_{R}^{2}+4\!\:\!\:\eta ^4\lambda ^{-8}\tilde{\varOmega}_6D_{R}^{2}-q,\\&\tau _{\theta \theta}=\tau _{\phi \phi}=2\!\:\lambda ^2\eta ^{-2}\tilde{\varOmega}_1+2\left( \eta ^2\lambda ^{-2}+\lambda ^4\eta ^{-4} \right) \tilde{\varOmega}_2-q.
	\end{split}
\end{align}

Upon inspecting Eq. \eqref{Eq13}, we identify the expression for the electric field, distinguished by its singular non-zero component, detailed as:
\begin{align}\label{Eq22}
	\begin{split}
		E_{r}=2\lambda ^2\eta ^{-4}\tilde{\varOmega}_4D_R+2\lambda ^{-2}\tilde{\varOmega}_5D_R+2\eta ^4\lambda ^{-6}\tilde{\varOmega}_6D_R.
	\end{split}
\end{align}

A scrutiny of Eqs. \eqref{Eq18} and \eqref{Eq20} reveals that all invariants (i.e., $I_1\left( \eta ,\lambda \right) $, $I_2\left( \eta ,\lambda \right) $, $K_4\left(\eta,D_R \right) $, $K_5\left( \lambda, D_R \right) $, and $K_6\left(\eta ,\lambda, D_R \right) $) exclusively hinge on three independent quantities: $\lambda$, pertaining to deformation; the factor $\eta$, in relation to finite growth; and the electric displacement $D_{R}$, indicative of electroelasticity. Given these findings, it is compelling to pursue a further refinement of the energy function $\tilde{\varOmega}$. The resulting streamlined energy function $\varOmega ^*\left(\eta ,\lambda ,D_R \right)$ is represented as:
\begin{align}\label{Eq23}
	\begin{split}
	\varOmega ^*\left(\eta ,\lambda ,D_R \right) &=\tilde{\varOmega}\left( I_1,I_2, K_4,K_5,K_6 \right) \\&=\tilde{\varOmega}\left( \eta ^4\lambda ^{-4}+2\eta ^{-2}\lambda ^2,2\!\:\eta ^2\!\:\lambda ^{-2}+\lambda ^4\eta ^{-4},D_{R}^{2}\eta ^{-4},D_{R}^{2}\lambda ^{-4},D_{R}^{2}\eta ^4\lambda ^{-8} \right).
	\end{split}
\end{align}
Given the above representation (note that $I_3=1$), we can elucidate the following derivatives:
\begin{align}\label{Eq24}
	\begin{split}
		&\varOmega _{\lambda}^{*}=\frac{\partial \varOmega ^*}{\partial \lambda}=4\left( \!\:\lambda \eta ^{-2}-\!\:\eta ^4\lambda ^{-5} \right) \tilde{\varOmega}_1+4\left( \lambda ^3\eta ^{-4}-\!\:\eta ^2\lambda ^{-3} \right) \tilde{\varOmega}_2-4\!\:D_{R}^{2}\lambda ^{-5}\tilde{\varOmega}_5-8\!\:D_{R}^{2}\!\:\eta ^4\lambda ^{-9}\tilde{\varOmega}_6,\\&\varOmega _{D_R}^{*}=\frac{\partial \varOmega ^*}{\partial D_R}=2\eta ^{-4}\tilde{\varOmega}_4D_R+2\lambda ^{-4}\tilde{\varOmega}_5D_R+2\eta ^4\lambda ^{-8}\tilde{\varOmega}_6D_R.
	\end{split}
\end{align}

Drawing from Eqs. \eqref{Eq21}, \eqref{Eq22} and \eqref{Eq24}, we establish the following pivotal and succinct relations:
\begin{align}\label{Eq25}
	\begin{split}
		\tau _{\theta \theta}-\tau _{rr}=\tau _{\phi \phi}-\tau _{rr}=\frac{\lambda}{2}\varOmega _{\lambda}^{*},\quad \mathrm{and}\quad E_r=\lambda ^2\varOmega _{D_R}^{*}.
	\end{split}
\end{align}

Furthermore, under the spherical symmetry assumption, the first Cauchy equation (see Eq. \eqref{Eq14}$_1$) for the considered deformation simplifies to ${{\partial \tau _{rr}}/{\partial r}}=r^{-1}\left( \tau _{\theta \theta}+\tau _{\phi \phi}-2\tau _{rr} \right) $.
Using this equation in conjunction with Eq. \eqref{Eq25}$_1$, we obtain the following important expression for determining the radial Cauchy stress:
\begin{align}\label{Eq26}
	\begin{split}
		\varOmega _{\lambda}^{*}\lambda r^{-1}=\frac{\mathrm{d}\tau _{rr}}{\mathrm{d}r}=\lambda r^{-1}\left( 1-J_{g}^{-1}\lambda ^3 \right) \frac{\mathrm{d}\tau _{rr}}{\mathrm{d}\lambda} \quad \Longrightarrow  \quad \mathrm{d}\tau _{rr}=\frac{\varOmega _{\lambda}^{*}}{1-J_{g}^{-1}\lambda ^3}\mathrm{d}\lambda .
	\end{split}
\end{align}

Note that the exterior electric field is assumed to be zero (the components of the Maxwell stress all vanish), there is no mechanical load at the outer boundary of the shell $r=r_o$, and on the inner boundary of the shell $r=r_i$, a pressure $P$ is applied (that is, $\tau _{rr}\left( r_i \right) =-P$). 
Integrating Eq. \eqref{Eq26} utilizing these boundary conditions derives the following expression:
\begin{align}\label{Eq27}
	\begin{split}
		P=\tau _{rr}\left( \lambda _o \right) -\tau _{rr}\left( \lambda _i \right) =\int_{\lambda _i}^{\lambda _o}{\frac{\varOmega _{\lambda}^{*}}{1-J_{g}^{-1}\lambda ^3}\mathrm{d}\lambda},
	\end{split}
\end{align}
and the components of the Cauchy stress tensor can be expressed as:
\begin{align}\label{Eq28}
	\begin{split}
		\tau _{rr}\left( \lambda \right) =\int_{\lambda _i}^{\lambda}{\frac{\varOmega _{\lambda}^{*}}{1-J_{g}^{-1}\lambda ^3}\mathrm{d}\lambda}-P, \quad \text{and} \quad \tau _{\theta \theta}=\tau _{\phi \phi}=\int_{\lambda _i}^{\lambda}{\frac{\varOmega _{\lambda}^{*}}{1-J_{g}^{-1}\lambda ^3}\mathrm{d}\lambda}+\frac{\lambda}{2}\varOmega _{\lambda}^{*}-P.
	\end{split}
\end{align}
where $\lambda_i={{r_i}/{R_i}}$ and $\lambda_o={{r_o}/{R_o}}$.

\subsection{Electroelastic growth base state}
Within the present framework, the relation \(\mathrm{curl}\mathbf{E}=\mathbf{0}\) is inherently satisfied. Simultaneously, the fundamental electroelastic equilibrium necessitates \(\mathrm{div}\mathbf{D}=0\). Accounting for the intricacies of spherical polar coordinates coupled with the characteristic spherically symmetric deformation, the equation elegantly simplifies to: $r^{-2}\mathrm{d}\left( r^2D_r \right) /\mathrm{d}r=0$, yielding $D_r=\lambda ^{-2}D_R={{c} /{r^2}}$, where \(c\) stands as an integration constant.
Further, recognizing the electric field as the negative gradient of the electric potential $\varPhi$, and drawing upon this relation, we can articulate the voltage $V$, which is defined as the potential difference between the inner $\varPhi_i$ and outer $\varPhi_o$ surfaces of the sphere. From the foregoing analysis, we arrive at the relation: $V=\int_{r_i}^{r_o}{\varepsilon ^{-1}D_r\mathrm{dr}}=c\varepsilon ^{-1}\left( r_{i}^{-1}-r_{o}^{-1} \right) $. Subsequently, the electric field and electric displacement are elucidated via the following formulation:
\begin{align}\label{Eq29}
	\begin{split}
		E_r=\frac{Vr_ir_o}{r^2\left( r_o-r_i \right)}, \quad  E_R=\frac{Vr_ir_oJ_g}{r^2\lambda ^2\left( r_o-r_i \right)},  \quad  D_r=\frac{\varepsilon Vr_ir_o}{r^2\left( r_o-r_i \right)}, \quad D_R=\frac{\varepsilon Vr_ir_o\lambda ^2}{r^2\left( r_o-r_i \right)}.
	\end{split}
\end{align}

To clarify and further elucidate the previously discussed results, we specify a particular energy function for illustration. The total Helmholtz free energy density \citep{dorfmann2014nonlinear} comprises the sum of the elastic component $\omega$ and the electric component $\omega^*$, i.e.,
\begin{align}\label{Eq30}
	\begin{split}
		\tilde{\varOmega}\left( \mathbf{A},\mathbf{D}_g \right) =\varOmega ^*=\omega\left( \mathbf{A} \right) +\omega^*\left( \mathbf{A},\mathbf{D}_g \right) =\omega\left( \mathbf{A} \right) +\frac{1}{2\varepsilon}\left( \alpha K_4+\beta K_5 \right).
	\end{split}
\end{align}
The term \(\omega(\mathbf{A})\) represents the intrinsic strain energy function pertinent to generalized materials. To succinctly integrate the influence of the electric biasing field, we introduce the supplemental term $\omega^*\left( \mathbf{A},\mathbf{D}_g \right) =\left( \alpha K_4+\beta K_5 \right) /2\varepsilon $.
This encapsulates two dimensionless material constants, \(\alpha\) and \(\beta\), which act as pivotal electroelastic coupling parameters.
Specifically, the coefficient \(\alpha\) remains decoupled from the stress manifestation. In contrast, the parameter \(\beta\) acts as a modulator, stiffening the material along the electric field orientation when juxtaposed against scenarios devoid of such a field. In a context where \(\alpha = 0\) and the strain energy function aligns with the behavior of a neo-Hookean material, the energy function mirrors the attributes of an ideal electroelastic material \citep{suo2008nonlinear}. Evidently, for a comprehensive mutual coupling, it is imperative to encompass both constants.

In advancing our discussion, we introduce several dimensionless parameters to facilitate subsequent analysis. These dimensionless quantities include:
\begin{align}\label{Eq31}
	\begin{split}
	\left[ \overline{\tilde{\varOmega}},\overline{\varOmega }^*,\overline{\omega},\overline{\omega}^*,\overline{P},\overline{\boldsymbol{\tau } } \right] =\frac{\left[ \tilde{\varOmega},\varOmega ^*,\omega,\omega^*,P,\boldsymbol{\tau } \right]}{\mu}, \quad \left[ \overline{D}_r,\overline{D}_R \right] =\frac{\left[ D_r,D_R \right]}{\sqrt{\varepsilon \mu}}, \quad \overline{V}=\frac{V\sqrt{\varepsilon \mu ^{-1}}}{H}=\frac{V\sqrt{\varepsilon \mu ^{-1}}}{R_o-R_i}.
	\end{split}
\end{align}
Herein, the material parameter \(\mu\) denotes the initial shear modulus.
It is imperative to accord special attention to several other dimensionless quantities delineated previously, such as $\lambda =r/R$, $\lambda _i=r_i/R_i$, $\lambda _o=r_o/R_o$, $\overline{r}_o=r_o/r_i$, and $\overline{R}_o=R_o/R_i$. 

Upon amalgamating Eqs. \eqref{Eq27}, \eqref{Eq29}, and \eqref{Eq30} we are led to the expression for the dimensionless voltage $\overline{V}$ as follows:
\begin{align}\label{Eq32}
	\begin{split}
		\overline{V}=\frac{\lambda _i\overline{\!\:r}_o}{\overline{R}_o-1}\sqrt{\frac{2}{\beta}\left( \frac{\overline{\!\:r}_o-1}{\overline{\!\:r}_{o}^{3}+\overline{\!\:r}_{o}^{2}+\overline{\!\:r}_o+1} \right) \left( \int_{\lambda _i}^{\lambda _o}{\overline{\omega}_{\lambda}\frac{J_g}{J_g-\lambda ^3}\mathrm{d}\lambda}-\overline{P} \right)}.
	\end{split}
\end{align}
from this formulation, it becomes evident that even if the pure strain energy function is devoid of the strain stiffening effect, the dimensionless voltage, attributed to the inclusion of the factor $\beta$, inherently exhibits strain stiffening characteristics. Concurrently, the Cauchy stress $\boldsymbol{\tau }$ also exhibits this peculiar characteristic. By assimilating Eqs. \eqref{Eq28} to \eqref{Eq31}, we deduce the dimensionless components of stress in the following manner:
\begin{align}\label{Eq33}
	\begin{split}
		&\overline{\tau }_{rr}=\frac{\overline{r}_{o}^{4}\left( \lambda ^4\left( \lambda _{i}^{3}-J_g \right) ^{{{4}/{3}}}-\lambda _{i}^{4}\left( \lambda ^3-J_g \right) ^{{{4}/{3}}} \right)}{\lambda ^4\!\:(\overline{r}_{o}^{4}-1)\left( \lambda _{i}^{3}-J_g \right) ^{{{4}/{3}}}}\left( \overline{P}-\int_{\lambda _i}^{\lambda _o}{\frac{\overline{\omega}_{\lambda}}{1-J_{g}^{-1}\lambda ^3}\mathrm{d}\lambda} \right) +\left( \int_{\lambda _i}^{\lambda}{\frac{\overline{\omega}_{\lambda}}{1-J_{g}^{-1}\lambda ^3}\mathrm{d}\lambda}-\overline{P} \right),   \\
		&\overline{\tau }_{\theta \theta}=\overline{\tau }_{\phi \phi}=\overline{\tau }_{rr}+\frac{\lambda}{2}\left( \overline{\omega}_{\lambda}-2\beta \lambda ^{-5}\overline{D}_{R}^{2} \right),
	\end{split}
\end{align}
and we express the term $\overline{D}_R$ as:
\begin{align}\label{Eq34}
	\begin{split}
		\overline{D}_R=-\frac{\left( \overline{R}_o-1 \right) \!\:\overline{R}_o\overline{\!\:V}\lambda _i\lambda _o\left( J_g-\lambda ^3 \right) ^{{{2}/{3}}}}{\left( J_g-\lambda _{i}^{3} \right) ^{{{2}/{3}}}\!\:\left( \lambda _i-\overline{R}_o\!\:\lambda _o \right)}.
	\end{split}
\end{align}

From the aforementioned equations, it is clear that the stress components embody three salient effects: the finite differential growth effect, the finite deformation effect, and the electroelastic coupling effect. 
Notably, the internal pressure within the spherical shell and electroelastic coupling parameters, such as $\beta$, play pivotal roles in influencing the distribution and temporal evolution of stress.
This, in turn, can induce symmetry-breaking in the morphology of accelerating healing biotissues. While the intrinsic stress naturally encompasses aspects of strain-stiffening effects, there is an enduring scientific curiosity to examine the repercussions of a solely strain-based energy function on the structural electroactive growth instability.

To gain deeper insights into the synergistic and coupling effects of electroelasticity and strain stiffening, we employ a dimensionless strain energy model, expressed in its most general form.
A prime example in this field is the Gent model, a renowned constitutive framework notable for its accuracy in depicting the significant deformation behaviors and the limited extensibility inherent in soft biological tissues.
The strain energy function for the Gent model is articulated as: $\overline{w}=\frac{w}{\mu}=-\frac{J_m}{2}\ln\mathrm{(}1-\frac{\mathbf{c}_a:\mathbf{I}-3}{J_m})$, where \( \mathbf{c}_a \) represents the right Cauchy-Green deformation tensor, and \( \mathbf{I} \) denotes the identity tensor. 
The parameter \( J_m \) epitomizes the level of strain stiffening, mirroring the singular boundary of the strain energy function. 
When the strain stiffening level \( J_m \) tends toward infinity, the Gent model converges to the neo-Hookean model. It is crucial to acknowledge that Eqs. \eqref{Eq32} and \eqref{Eq33} are universally applicable to all the formulated strain energy models. Essentially, they embody the generic forms of dimensionless voltage and dimensionless Cauchy stress in the electroactive growth foundational state.
By incorporating the Gent strain energy function into the specified equations (i.e., Eqs. \eqref{Eq32} and \eqref{Eq33}), we initiate a comprehensive investigation of the foundational state of a growth spherical shell, regulated by electroelasticity. We place a special focus on its strain-stiffening characteristics ($\beta$ and $J_m$) and the implications of differential growth patterns, such as isotropic growth ($\gamma=\eta\ne1$), area growth ($\gamma=1$, $\eta\ne1$), and fiber growth ($\gamma\ne1$, $\eta=1$).

\begin{figure}[t!]
	\begin{center}
		\includegraphics[width=1\textwidth]{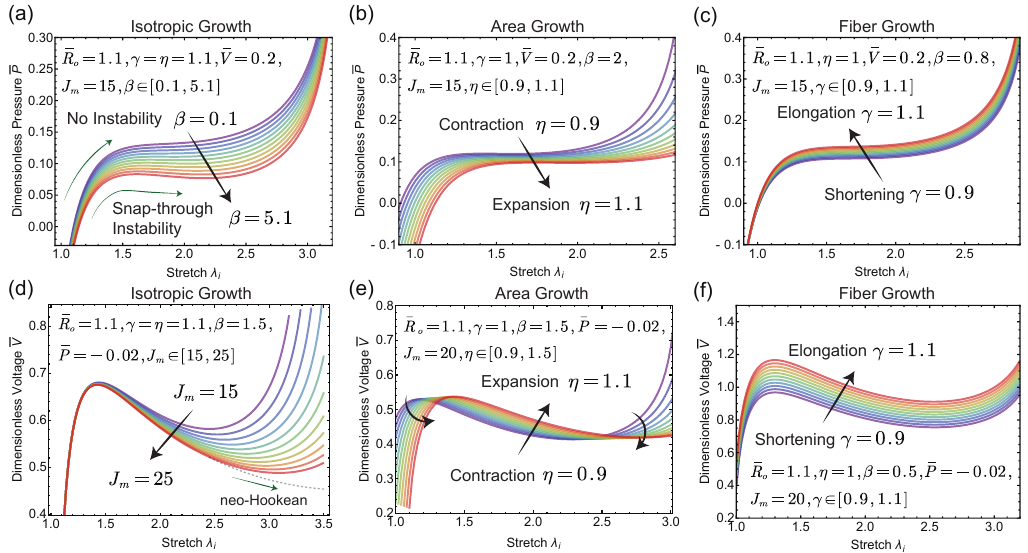}
		\caption{The impact of strain-stiffening effects on dimensionless internal pressure and voltage in spherically growing shells subjected to differential growth.
		(a) Internal pressure as a function of the stretch $\lambda_{i}$ at the inner surface under varying electroelastic coupling parameters $\beta$. 
		(b) Internal pressure-stretch curve during area growth ($\gamma=1$, $\eta\ne1$).
		(c) Internal pressure-stretch curve during fiber growth ($\gamma\ne1$, $\eta=1$).
		(d) Variation of dimensionless voltage as a function of the stretch $\lambda_{i}$ at the inner surface under different levels of strain stiffening $J_m$.
		(e) Dimensionless voltage as a function of the inner surface stretch under conditions of area growth.
		(f) Dimensionless voltage as a function of the inner surface stretch in the context of fiber growth.
		}\label{Fig-P&V}
	\end{center}
\end{figure}

Fig. \ref{Fig-P&V} delineates the evolution of dimensionless internal pressure $\bar{P}$ and dimensionless voltage $\bar{V}$ in relation to the inner surface stretch $\lambda_{i}$, parameterized by varying strain stiffening modulators ($J_m$ and $\beta$). A salient inference can be drawn: as $\bar{P}$ and $\bar{V}$ increase, all curves manifest a bifurcation behavior typified by limit-point instability. This is conspicuously evidenced by non-monotonic behavior in the curves, characterized by local extrema.
In the regime of strain-stiffening soft materials, this limit-point instability is immediately succeeded by a phenomenon known as an inflation jump, a specific form of snap-through instability. Within the framework of pressure-stretch or voltage-stretch response curves, this is typified by a local maximum followed by a subsequent local minimum within a finite stretch regime.
Fig. \ref{Fig-P&V}(a) reveals that varying the electroelastic coupling parameter $\beta$, within the span of 0.1 to 5.1, can elicit the occurrence of snap-through instability. For this illustration, the isotropic growth is maintained with a growth factor $\gamma = \eta = 1.1$. Contrarily, Figs. \ref{Fig-P&V}(b) and \ref{Fig-P&V}(c), which correspond to area and fiber growths respectively, do not display snap-through instability. Instead, a modulatory effect on internal pressure is observed, characterized by contraction and expansion ($\eta$ varies from 0.9 to 1.1) or by shortening and elongation ($\gamma$ ranges from 0.9 to 1.1).
Finally, Fig. \ref{Fig-P&V}(d) elucidates that at a constant internal pressure (e.g., $\bar{P} = -0.02$), the strain stiffening level $J_m$ acts as a significant modulator for the snap-through instability response curve. Concurrently, an increase in $J_m$ results in a decrement in the dimensionless voltage $\bar{V}$. In asymptotic conditions, as $J_m$ approaches infinity, the constitutive behavior of biological tissue degenerates to a neo-Hookean model, wherein only limit-point instability prevails, devoid of any inflation jump or snap-through instability. In the context of anisotropic growth patterns, depicted in Figs. \ref{Fig-P&V}(e) and \ref{Fig-P&V}(f), a marked influence on the dimensionless voltage is discernible, manifesting as hysteresis behavior in all response curves.

\begin{figure}[t!]
	\begin{center}
		\includegraphics[width=1\textwidth]{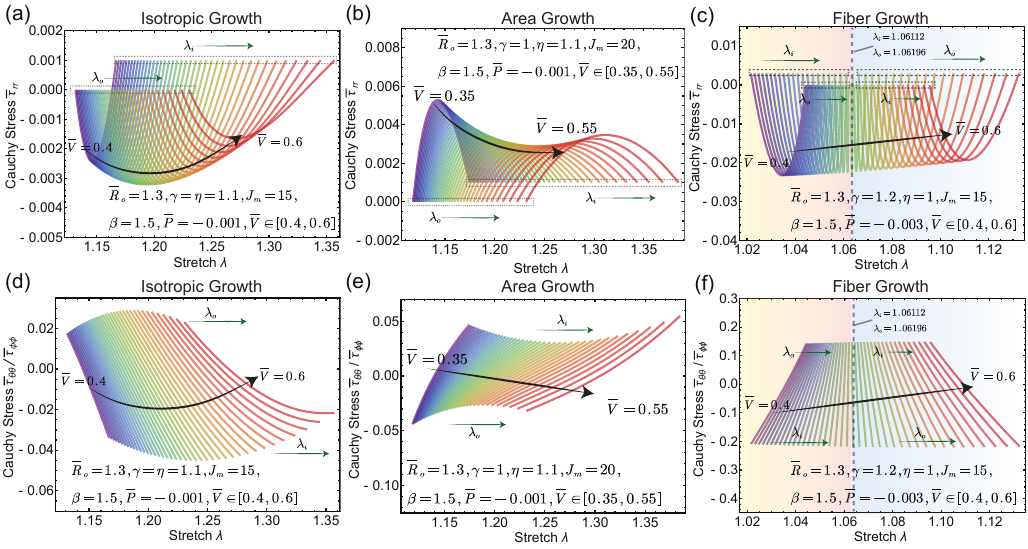}
		\caption{
		Significant impacts of differential growth and dimensionless voltage on Cauchy stress under strain stiffening.
		Radial Cauchy stress as a function of stretch under (a) isotropic growth, (b) area growth, and (c) fiber growth conditions.
		Evolution of Cauchy stress in polar and azimuthal directions as a function of stretch under (d) isotropic growth, (e) area growth, and (f) fiber growth conditions. It is noteworthy that due to the geometrical characteristics of the spherical shell, the Cauchy stresses in the polar and azimuthal directions are equal.
						}\label{Fig-cauchy}
	\end{center}
\end{figure}

Fig. \ref{Fig-cauchy} presents the variation of Cauchy stress components ($\overline{\tau}_{rr}$, $\overline{\tau}_{\theta\theta}$, and $\overline{\tau}_{\phi\phi}$) as functions of the stretch $\lambda$, under different modes of material growth: isotropic ($\gamma=\eta=1.1$), area-based ($\gamma=1, \eta=1.1$), and fiber-oriented ($\gamma=1.2, \eta=1$).
In the regime of isotropic growth, as illustrated in Fig. \ref{Fig-cauchy}(a) and Fig. \ref{Fig-cauchy}(d), the entire family of Cauchy stress curves translates rightward with the escalation in dimensionless voltage $\bar{V}$, ranging between 0.4 and 0.6. Accompanied by a non-zero internal pressure $\bar{P}$, the radial Cauchy stress initially decreases, reaches a local minimum, and subsequently ascends, as evinced in Fig. \ref{Fig-cauchy}(a). Concurrently, the Cauchy stresses in both polar and azimuthal directions undergo a monotonic descent, as exhibited in Fig. \ref{Fig-cauchy}(d).
For area-based growth, Fig. \ref{Fig-cauchy}(b) indicates that the radial Cauchy stress initially escalates with an increasing stretch, attaining a peak value before regressing to a designated internal pressure level.
Here, it is imperative to differentiate between the inner surface stretch $\lambda_i$ and the outer surface stretch $\lambda_o$. Fig. \ref{Fig-cauchy}(e) portrays the Cauchy stresses in polar and azimuthal orientations as monotonically increasing. 
In addition, variations in dimensionless voltage $\bar{V}$ induce a rightward shift in all Cauchy stress curves.
In the context of fiber growth, the Cauchy stress components in the polar and azimuthal orientations exhibit bifurcated behaviors contingent upon whether the stretch ratio is below or exceeds 1.0611. 
For $\lambda < 1.0611$, these stress components manifest a monotonous ascension, whereas for $\lambda > 1.0611$, a monotonous descent is observed. 
Regarding the radial Cauchy stress, its global evolution features a decline followed by a surge; however, the directional nuances of this trajectory diverge from those observed in previous growth modalities. For $\lambda < 1.0611$, the radial Cauchy stress progresses from the inner surface stretch $\lambda_i$ towards the outer surface stretch $\lambda_o$. Contrarily, for $\lambda > 1.0611$, this directional trend is inversed.

\begin{figure}[t!]
	\begin{center}
		\includegraphics[width=1\textwidth]{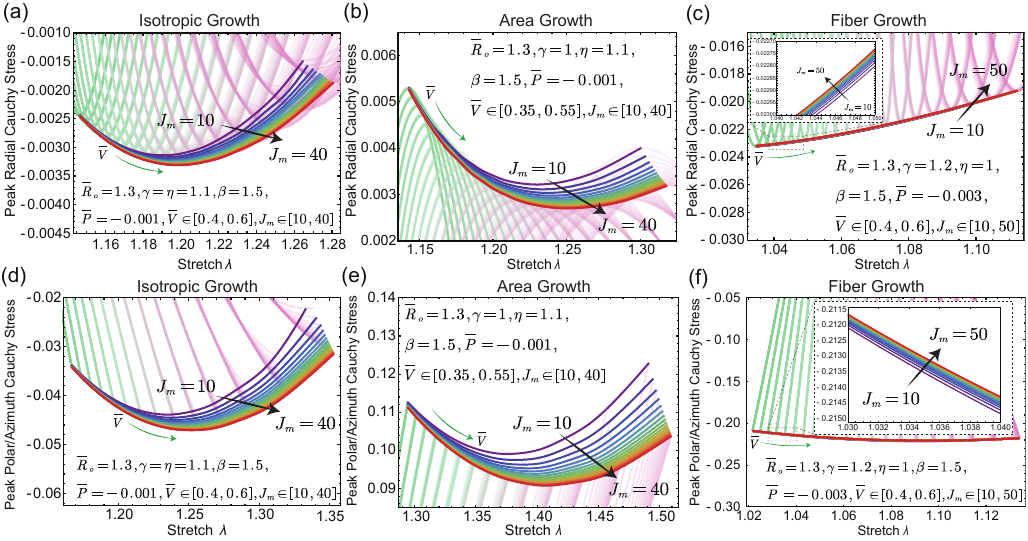}
		\caption{
		Effects of strain stiffening on peak Cauchy stresses: Characterization of peak radial Cauchy stress as a function of stretch is conducted under different growth conditions; specifically, (a) isotropic growth with a strain-stiffening level $J_m$ ranging from 10 to 40, (b) area growth with $J_m$ ranging from 10 to 40, and (c) fiber growth with $J_m$ ranging from 10 to 50. Similarly, the evolution of peak Cauchy stress in the polar and azimuthal directions is analyzed as a function of stretch under (d) isotropic growth with $J_m$ ranging from 10 to 40, (e) area growth with $J_m$ ranging from 10 to 40, and (f) fiber growth with $J_m$ ranging from 10 to 50.
				}\label{Fig-Peak}
	\end{center}
\end{figure}

To further investigate the impact of strain stiffening on electroactive differential growth, we present the evolution of peak stress as a function of stretch in Fig. \ref{Fig-Peak}.
For isotropic (see Figs. \ref{Fig-Peak}a and \ref{Fig-Peak}d) and area-dependent growth (see (see Figs. \ref{Fig-Peak}b and \ref{Fig-Peak}e)), the peak radial and polar/azimuthal Cauchy stresses exhibit parabolic nonlinear evolution. 
More importantly, as the level of strain stiffening increases, for instance from 10 to 40, these peak stresses consistently diminish. 
Conversely, in the case of fiber growth, the peak radial Cauchy stress exhibits a monotonically increasing trend (see Fig. \ref{Fig-Peak}c), while the polar/azimuthal Cauchy stresses remain relatively constant (see Fig. \ref{Fig-Peak}f, for details). 
Intriguingly, as the level of strain stiffening escalates, all three types of peak Cauchy stresses increase within a relatively confined range.

\section{Perturbation analysis of electroelastic differential growth}\label{section4}
\subsection{Incremental equilibrium and Maxwell's Equations}
In the present study, we concentrate on the governing equations for linearized incremental deformations and superimposed electric fields within a pre-established, finitely deformed configuration possessing a well-defined electric field. Utilizing Lagrangian coordinates as the foundational framework, we formulate the generalized incremental equations and their corresponding push-forward representations. Moreover, we rigorously derive the constitutive moduli tensors, elucidating the intricate coupling mechanisms between incremental mechanical and electric phenomena. 
By introducing an incremental motion function $\chi ^{\left( 1 \right)}\left( \mathbf{x} \right) $, we establish a first-order kinematic relationship $\tilde{\chi}=\chi \left( \mathbf{X} \right) +\epsilon \chi ^{\left( 1 \right)}\left( \mathbf{x} \right) $, where $\epsilon$ represents a small perturbation parameter. It is noteworthy that the incremental motion function $\chi ^{\left( 1 \right)}\left( \mathbf{x} \right) $ is a function of the current position vector $\mathbf{x}$.
Consequently, upon the introduction of $\mathbf{F}^{(1)} = \mathrm{grad} \left( \chi^{(1)}(\mathbf{x}) \right)$ and $\mathring{\mathbf{F}} = \mathbf{F}^{(1)} \mathbf{F}$, and by executing the gradient operation in the context of the Lagrangian description on both sides of the established relationship, we derive the ensuing expressions:
\begin{align}\label{Eq35}
	\begin{split}
	\tilde{\mathbf{F}}=\mathrm{Grad}\left( \tilde{\chi} \right) =\mathrm{Grad}\left( \chi \left( \mathbf{X} \right) \right) +\epsilon \mathrm{Grad}\left( \chi ^{\left( 1 \right)}\left( \mathbf{x} \right) \right) =\mathbf{F}+\epsilon \mathrm{grad}\left( \chi ^{\left( 1 \right)}\left( \mathbf{x} \right) \right) \mathbf{F}=\mathbf{F}+\epsilon \mathring{\mathbf{F}}.
	\end{split}
\end{align}

Subsequently, by conducting a first-order expansion on the incremental pure elastic deformation tensor $\tilde{\mathbf{A}}$, we arrive at the relationship $\tilde{\mathbf{A}} = \mathbf{A} + \epsilon \mathring{\mathbf{A}}$, where $\mathring{\mathbf{A}} = \mathbf{A}^{(1)} \mathbf{A}$. 
It is noteworthy that, on one hand, $\mathring{\mathbf{F}} = \mathbf{F}^{(1)} \mathbf{F} = \mathbf{F}^{(1)} \mathbf{A} \mathbf{G}$, and on the other hand, the relationship $\mathring{\mathbf{F}} = \mathring{\mathbf{A}} \mathbf{G}$ holds true. Consequently, we establish the following relationship:
\begin{align}\label{Eq36}
	\begin{split}
		\mathring{\mathbf{A}}=\mathbf{A}^{\left( 1 \right)}\mathbf{A}=\mathbf{F}^{\left( 1 \right)}\mathbf{A}  \quad \Longrightarrow  \quad \mathbf{A}^{\left( 1 \right)}=\mathbf{F}^{\left( 1 \right)}.
	\end{split}
\end{align}

Furthermore, under the incompressibility constraint, the isochoric condition necessitates that the Jacobian determinant of the pure elastic deformation tensor be unity, i.e., $\det ( \tilde{\mathbf{A}} ) =\det \left( ( \mathbf{I}+\epsilon \mathbf{F}^{\left( 1 \right)} ) \mathbf{A} \right) =1$. 
As a corollary, with respect to the incremental incompressibility constraint, we derive the following critical expression:
\begin{align}\label{Eq37}
	\begin{split}
	\det \left( \mathbf{I}+\epsilon \mathbf{F}^{\left( 1 \right)} \right) =1+\epsilon \mathrm{tr}\left( \mathbf{F}^{\left( 1 \right)} \right) =1 \quad \Longrightarrow \quad \mathrm{tr}\left( \mathbf{F}^{\left( 1 \right)} \right) =0.
	\end{split}
\end{align}

To formulate the incremental equations, we commence by adopting the Lagrangian description (see Eq. \eqref{Eq7} and Eq. \eqref{Eq9}, for details) to define the incremental forms of the nominal stress tensor $\mathbf{S}^{\left( 1 \right)}$ and the Lagrangian electric field $\mathbf{E}_{l}^{\left( 1 \right)}$. Utilizing the principle of virtual work, we consequently arrive at:
\begin{align}\label{Eq38}
	\begin{split}
		\mathbf{S}^{\left( 1 \right)}=J\mathbf{G}^{-1}\delta \left( \frac{\partial \tilde{\varOmega}\left( \mathbf{A},\mathbf{D}_g \right)}{\partial \mathbf{A}}-q\mathbf{A}^{-1} \right) =J\mathbf{G}^{-1}\left( \frac{\partial ^2\tilde{\varOmega}}{\partial \mathbf{A}\partial \mathbf{A}}\mathbf{F}^{(1)}\mathbf{A}+\frac{\partial ^2\tilde{\varOmega}}{\partial \mathbf{A}\partial \mathbf{D}_g}\mathbf{D}_{g}^{(1)}-q^{(1)}\mathbf{A}^{-1}+q\mathbf{A}^{-1}\mathbf{F}^{(1)} \right),
	\end{split}
\end{align}
and
\begin{align}\label{Eq39}
	\begin{split}
		\mathbf{E}_{l}^{\left( 1 \right)}=\mathbf{G}^{\mathrm{T}}\delta \left( \frac{\partial \tilde{\varOmega}\left( \mathbf{A},\mathbf{D}_g \right)}{\partial \mathbf{D}_g} \right) =\mathbf{G}^{\mathrm{T}}\left( \frac{\partial ^2\tilde{\varOmega}}{\partial \mathbf{D}_g\partial \mathbf{A}}\mathbf{F}^{\left( 1 \right)}\mathbf{A}+\frac{\partial ^2\tilde{\varOmega}}{\partial \mathbf{D}_g\partial \mathbf{D}_g}\mathbf{D}_{g}^{\left( 1 \right)} \right).
	\end{split}
\end{align}

Upon introducing the fourth-order $\mathbf{\mathcal{A}} ^*=\partial ^2\tilde{\varOmega}/\partial \mathbf{A}\partial \mathbf{A}$, third-order $\boldsymbol{\varGamma }^*={{\partial ^2\tilde{\varOmega}}/{\partial \mathbf{A}\partial \mathbf{D}_g}}$, and second-order $\mathbf{\mathcal{K} } ^*={{\partial ^2\tilde{\varOmega}}/{\partial \mathbf{D}_g\partial \mathbf{D}_g}}$ electroelastic moduli tensors, we derive the incremental forms of both the nominal stress tensor and the Lagrangian electric field as:
\begin{align}\label{Eq40}
	\begin{split}
		\mathbf{S}^{\left( 1 \right)}=J\mathbf{G}^{-1}\left( \mathbf{\mathcal{A}} ^*\mathbf{F}^{(1)}\mathbf{A}+\boldsymbol{\varGamma }^*\mathbf{D}_{g}^{(1)}-q^{(1)}\mathbf{A}^{-1}+q\mathbf{A}^{-1}\mathbf{F}^{(1)} \right),
	\end{split}
\end{align}
and
\begin{align}\label{Eq41}
	\begin{split}
		\mathbf{E}_{l}^{\left( 1 \right)}=\mathbf{G}^{\mathrm{T}}\left( \boldsymbol{\varGamma }^*\mathbf{F}^{\left( 1 \right)}\mathbf{A}+\mathbf{\mathcal{K}} ^*\mathbf{D}_{g}^{\left( 1 \right)} \right).
	\end{split}
\end{align}

Given that the incremental deformations are superimposed upon the current state, subsequent to obtaining the incremental forms of the nominal stress tensor and the Lagrangian electric field, it becomes imperative to consider their push-forward representations, denoted as $\mathbf{S}_{0}^{\left( 1 \right)}$ and $\mathbf{E}_{l0}^{\left( 1 \right)}$. Based on the relationships of $\mathbf{S}_{0}^{\left( 1 \right)}=J^{-1}\mathbf{F}\mathbf{S}^{\left( 1 \right)}$ and $\mathbf{E}_{l0}^{\left( 1 \right)}=\mathbf{F}^{-\mathrm{T}}\mathbf{E}_{l}^{\left( 1 \right)}$, we arrive at the following pivotal equations:
\begin{align}\label{Eq42}
	\begin{split}
	\mathbf{S}_{0}^{\left( 1 \right)}=\mathbf{\mathcal{A}} _{0}^{*}\mathbf{F}^{(1)}+\boldsymbol{\varGamma }_{0}^{*}\mathbf{D}_{g0}^{(1)}-q^{(1)}\mathbf{I}+q\mathbf{F}^{(1)},
	\end{split}
\end{align}
and
\begin{align}\label{Eq43}
	\begin{split}
		\mathbf{E}_{l0}^{\left( 1 \right)}=\left( \boldsymbol{\varGamma }_{0}^{*} \right) ^{\mathrm{T}}\mathbf{F}^{\left( 1 \right)}+\mathbf{\mathcal{K}} _{0}^{*}\mathbf{D}_{g0}^{(1)}.
	\end{split}
\end{align}
where $\mathcal{A} _{0jilk}^{*}=F_{j\alpha}F_{l\beta}\mathcal{A} _{\alpha i\beta k}^{*}$,  $\varGamma _{0jik}^{*}=F_{j\alpha}F_{\beta k}^{-1}\varGamma _{\alpha i\beta}^{*}$, and $ \mathcal{K} _{0ij}^{*}=F_{\alpha i}^{-1}F_{\beta j}^{-1}\mathcal{K} _{\alpha \beta}^{*}$ represent the transformation relationships for the electroelastic moduli.

Finally, the updated versions of the incremental equilibrium and Maxwell's equations can be articulated as follows:
\begin{align}\label{Eq44}
	\begin{split}
	\mathrm{div}\left( \mathbf{S}_{0}^{\left( 1 \right)} \right) =\mathbf{0}, \quad \mathrm{div}\left( \mathbf{D}_{g0}^{(1)} \right) =\mathrm{tr}\left( \mathrm{grad}\left( \mathbf{D}_{g0}^{(1)} \right) \right) =0, \quad \text{and} \quad \mathrm{curl}\left( \mathbf{E}_{l0}^{\left( 1 \right)} \right) =\mathbf{0}.
	\end{split}
\end{align}

It is worth noting that upon introducing the incremental electric potential, denoted as $\varPhi^{(1)}$, we arrive at the equation $\mathbf{E}_{l0}^{(1)} = -\mathrm{grad}(\varPhi^{(1)})$. This naturally leads to the conclusion that the equation $\mathrm{curl}(\mathbf{E}_{l0}^{(1)}) = \mathbf{0}$ is inherently satisfied.

\subsection{Incremental field equations and Stroh resolution framework}
Here, we explore a spherical shell exhibiting differential growth, alongside electroactive and strain-stiffening characteristics.
We describe the first-order generalized deformation, triggered by bifurcation, as $\chi ^{\left( 1 \right)}=\left[ u,v,w \right] ^{\mathrm{T}}$, where $u$, $v$, and $w$ are functions of the standard spherical coordinates $\left( r,\theta ,\phi \right) $.
In the conventional methodology, \( \chi^{(1)} \) is typically expanded in terms of spherical harmonics, subsequently leading to the solution of the associated incremental equations. 
However, when determining the parameter values at which bifurcation occurs, some studies have shown that the displacement field in the azimuthal direction can be decoupled during the resolution of these incremental equations \citep{wang1972stability, amar2005growth}. This means that the derived system of differential equations is independent of \(w\), indicating that \(w\) has no impact on the perturbation analysis \citep{ciarletta2013buckling}. 
To streamline computations, it is judicious to postulate the incremental displacement field (axisymmetric deformations) in the following form: 
\begin{align}\label{Eq45}
	\begin{split}
		\chi ^{\left( 1 \right)}=\left[ u, v ,0 \right] ^{\mathrm{T}}.
	\end{split}
\end{align}
where \( u \) and \( v \) are independent of \( \phi \).
Upon this formulation, the incremental displacement gradient can subsequently be determined as follows:
\begin{align}\label{Eq46}
	\begin{split}
		\mathbf{F}^{\left( 1 \right)}=\mathrm{grad}\left( \chi ^{\left( 1 \right)} \right) =\left[ \begin{matrix}	\frac{\partial u}{\partial r}&		r^{-1}\left( \frac{\partial u}{\partial \theta}-v \right)&		0\\	\frac{\partial v}{\partial r}&		r^{-1}\left( \frac{\partial v}{\partial \theta}+u \right)&		0\\	0&		0&		\frac{u\sin \theta +v\cos \theta}{r\sin \theta}\\\end{matrix} \right].
	\end{split}
\end{align}

Evidently, under the stipulated geometric constraints given by Eq. \eqref{Eq37}, we derive the following relationship:
\begin{align}\label{Eq47}
	\begin{split}
		\mathrm{tr}\left( \mathbf{F}^{\left( 1 \right)} \right) ={ \frac{\partial u}{\partial r}+r^{-1}\left( \frac{\partial v}{\partial \theta}+u \right) +r^{-1}\left( u+v\cot \theta \right) =0}.
	\end{split}
\end{align}

To conform to the constraints set forth by Eq. \eqref{Eq44}, the technique of variable separation can be invoked to establish the incremental mechanics deformation, as well as the associated electroelastic deformation field, as functions represented by spherical harmonics. Without loss of generality, the following ansatz is proposed:
\begin{align}\label{Eq48}
	\begin{split}
		&\left[ u,S_{0rr}^{\left( 1 \right)},\varPhi ^{\left( 1 \right)},D_{g0r}^{(1)},q^{\left( 1 \right)} \right] =\left[ \mathbb{U} \left( r \right) ,\mathbb{S} _{rr}\left( r \right) ,\varXi \left( r \right) ,\mathbb{D} _r\left( r \right) ,Q\left( r \right) \right] \mathbb{P} _m\left( \cos \theta \right) ,\\&\left[ v,S_{0r\theta}^{\left( 1 \right)} \right] =\frac{1}{\sqrt{m\left( m+1 \right)}}\left[ \mathbb{V} \left( r \right) ,\mathbb{S} _{r\theta}\left( r \right) \right] \frac{\mathrm{d}\mathbb{P} _m\left( \cos \theta \right)}{\mathrm{d}\theta},
	\end{split}
\end{align}
where $\mathbb{P} _m\left( x \right) $ denotes the Legendre polynomial of order \( m \), and $\mathbb{P} _m\left( \cos \theta \right) $ satisfies the following equation:
\begin{align}\label{Eq49}
	\begin{split}
		\frac{\mathrm{d}^2\mathbb{P} _m(\cos \theta )}{\mathrm{d}\theta ^2}+\cot \theta \frac{\mathrm{d}\mathbb{P} _m(\cos \theta )}{\mathrm{d}\theta}+m\left( m+1 \right) \mathbb{P} _m(\cos \theta )=0.
	\end{split}
\end{align}

Due to the inherent satisfaction of Eq. \eqref{Eq44}$_3$, the residual incremental field equations, specifically the incremental equations of mechanics equilibrium and the incremental Maxwell's equations, can be elegantly expressed in the following refined formulation:
\begin{align}\label{Eq50}
	\begin{split}
	&\frac{\partial S_{0rr}^{\left( 1 \right)}}{\partial r}+\frac{1}{r}\left( \frac{\partial S_{0\theta r}^{\left( 1 \right)}}{\partial \theta}+2S_{0rr}^{\left( 1 \right)}-S_{0\theta \theta}^{\left( 1 \right)}-S_{0\phi \phi}^{\left( 1 \right)}+S_{0\theta r}^{\left( 1 \right)}\cot \theta \right) =0,   \\	&\frac{\partial S_{0r\theta}^{\left( 1 \right)}}{\partial r}+\frac{1}{r}\left( \frac{\partial S_{0\theta \theta}^{\left( 1 \right)}}{\partial \theta}+2S_{0r\theta}^{\left( 1 \right)}+S_{0\theta r}^{\left( 1 \right)}+\left( S_{0\theta \theta}^{\left( 1 \right)}-S_{0\phi \phi}^{\left( 1 \right)} \right) \cot \theta \right) =0,
	\end{split}
\end{align}
and
\begin{align}\label{Eq51}
	\begin{split}
	\frac{1}{r^2}\frac{\partial \left( r^2D_{g0r}^{(1)} \right)}{\partial r}+\frac{1}{r}\frac{\partial D_{g0\theta}^{(1)}}{\partial \theta}+\frac{\cot \theta}{r}D_{g0\theta}^{(1)}=0.
	\end{split}
\end{align}

At this stage, through the integration of the displacement vector \(\mathcal{U} \left( r \right) =\left[ \mathbb{U} \left( r \right) ,\mathbb{V} \left( r \right) ,\left( r\mathbb{D} _r\left( r \right) \right) \right] ^{\mathrm{T}}\) and the traction vector \(\mathcal{S} \left( r \right) =\left[ \left( r\mathbb{S} _{rr}\left( r \right) \right) ,\left( r\mathbb{S} _{r\theta}\left( r \right) \right) ,\varXi \left( r \right) \right] ^{\mathrm{T}}\), we formulate an incremental electroelastic Stroh vector \(\mathcal{Y} \left( r \right) =\left[ \mathcal{U} \left( r \right) ,\mathcal{S} \left( r \right) \right] ^{\mathrm{T}}\). This enables the transformation of the preceding system of partial differential equations, specifically spanning Eqs. \eqref{Eq47} to \eqref{Eq51}, into a set of first-order vectorial ordinary differential equations. This transformation is colloquially known as the Stroh formalism, yielding the following equations:
\begin{align}\label{Eq52}
	\begin{split}
		\frac{\mathrm{d}}{\mathrm{d}r}\mathcal{Y} \left( r \right) =r^{-1}\mathcal{G} \left( r \right) \mathcal{Y} \left( r \right) =r^{-1}\left[ \begin{matrix}	\mathcal{G} _{11}&		\mathcal{G} _{12}\\	\mathcal{G} _{21}&		\mathcal{G} _{22}\\\end{matrix} \right] \mathcal{Y} \left( r \right),
	\end{split}
\end{align}
whrere $\mathcal{G}$ represents the Stroh matrix. 
Subsequent to this, a functional relationship $\mathcal{S} \left( r \right) =\mathcal{Z} ^i\left( r,r_i \right) \mathcal{U} \left( r \right) $ is established between the incremental displacement vector and the incremental traction vector, characterized by the inner surface impedance matrix $\mathcal{Z} ^i\left( r,r_i \right)$. By integrating $\mathcal{Z} ^i\left( r,r_i \right)$ into Eq. \eqref{Eq52}, we elucidate the ensuing pair of vectorial ordinary differential equations:
\begin{align}\label{Eq53}
	\begin{split}
	&\frac{\mathrm{d}\mathcal{U} \left( r \right)}{\mathrm{d}r}=r^{-1}\left( \mathcal{G} _{11}\mathcal{U} \left( r \right) +\mathcal{G} _{12}\mathcal{Z} ^i\left( r,r_i \right) \mathcal{U} \left( r \right) \right) ,\\	&\frac{\mathrm{d}\mathcal{Z} ^i\left( r,r_i \right)}{\mathrm{d}r}\mathcal{U} \left( r \right) +\frac{\mathrm{d}\mathcal{U} \left( r \right)}{\mathrm{d}r}\mathcal{Z} ^i\left( r,r_i \right) =r^{-1}\left( \mathcal{G} _{21}\mathcal{U} \left( r \right) +\mathcal{G} _{22}\mathcal{Z} ^i\left( r,r_i \right) \mathcal{U} \left( r \right) \right).
	\end{split}
\end{align}
Upon close examination of Eq. \eqref{Eq53} and the elimination of the incremental displacement vector, we arrive at a formulation that is inherently characterized by a Riccati differential equation.
Following this, through the execution of a mathematical transformation on the equation, we obtain its dimensionless form, articulated as:
\begin{align}\label{Eq54}
	\begin{split}
	\frac{\mathrm{d}\overline{\mathcal{Z} }^i\left( \lambda ,\lambda _i \right)}{\mathrm{d}\lambda}=\frac{J_g}{\lambda \left( J_g-\lambda ^3 \right)}\left( \overline{\mathcal{G} }_{21}+\overline{\mathcal{G} }_{22}\overline{\mathcal{Z} }^i\left( \lambda ,\lambda _i \right) -\overline{\mathcal{Z} }^i\left( \lambda ,\lambda _i \right) \overline{\mathcal{G} }_{11}-\overline{\mathcal{Z} }^i\left( \lambda ,\lambda _i \right) \overline{\mathcal{G} }_{12}\overline{\mathcal{Z} }^i\left( \lambda ,\lambda _i \right) \right) .
	\end{split}
\end{align}
where the sub-matrix configuration of the dimensionless Stroh matrix is elaborated in Eqs.~\eqref{Eq A1} and \eqref{Eq A2}.

To elucidate the effects of strain stiffening and electroelastic coupling, herein we present the refined formulations for three types of dimensionless electroelastic moduli tensors:
\begin{align}\label{Eq57}
	\begin{split}
	&\overline{\mathcal{A} }_{0piqj}^{*}=4\overline{\tilde{\Omega}}_{11}\left( \mathbf{b}_a \right) _{ip}\left( \mathbf{b}_a \right) _{jq}+2\overline{\tilde{\Omega}}_1\delta _{ij}\left( \mathbf{b}_a \right) _{pq}+2\overline{\tilde{\Omega}}_5\delta _{ij}\bar{D}_p\bar{D}_q,\\
	&\overline{\varGamma }_{0piq}^{*}=2\overline{\tilde{\Omega}}_5\left( \delta _{pq}\bar{D}_i+\delta _{iq}\bar{D}_p \right) , \quad \overline{\mathcal{K} }_{0ij}^{*}=2\left( \overline{\tilde{\Omega}}_4\left( \mathbf{b}_a \right) _{ij}^{-1}+\overline{\tilde{\Omega}}_5\delta _{ij} \right) .
	\end{split}
\end{align}
	
Given the imperative to account for the influence of internal pressure, the dimensionless incremental initial conditions can be reformulated as follows:
\begin{align}\label{Eq58}
	\begin{split}
		\overline{\mathcal{Z} }^i\left( \lambda _i,\lambda _i \right) =\left[ \begin{matrix}	-2\overline{P}&		\overline{P}\kappa&		0\\	\overline{P}\kappa&		-\overline{P}&		0\\	0&		0&		0\\\end{matrix} \right].
	\end{split}
\end{align}
Initiating from these prescribed conditions and by solving Eq. \eqref{Eq54}, we uncover nontrivial solutions and critical differential growth conditions that give rise to electroactive instability states when $\overline{\mathcal{Z} }^i\left( \lambda _o,\lambda _i \right)$ adheres to the subsequent relationship:
\begin{align}\label{Eq59}
	\begin{split}
	\det \left( \overline{\mathcal{Z} }^i\left( \lambda _o,\lambda _i \right) \right) =0.
	\end{split}
\end{align}
Upon achieving the convergence criterion set by Eq. \eqref{Eq59}, we can ascertain the ratio between the incremental displacements as follows:
\begin{align}\label{Eq60}
	\begin{split}
	\overline{\mathbb{K} }_{uv}=\frac{\overline{\mathbb{U} }\left( \lambda _o \right)}{\overline{\mathbb{V} }\left( \lambda _o \right)}=-\frac{\left( \overline{\mathcal{Z} }_{12}^{i}\left( \lambda _o,\lambda _i \right) \overline{\mathcal{Z} }_{33}^{i}\left( \lambda _o,\lambda _i \right) -\overline{\mathcal{Z} }_{13}^{i}\left( \lambda _o,\lambda _i \right) \mathcal{Z} _{32}^{i}\left( \lambda _o,\lambda _i \right) \right)}{\left( \overline{\mathcal{Z} }_{11}^{i}\left( \lambda _o,\lambda _i \right) \overline{\mathcal{Z} }_{33}^{i}\left( \lambda _o,\lambda _i \right) -\overline{\mathcal{Z} }_{13}^{i}\left( \lambda _o,\lambda _i \right) \mathcal{Z} _{31}^{i}\left( \lambda _o,\lambda _i \right) \right)}.
	\end{split}
\end{align}
Subsequently, to delineate the through-thickness distribution of the incremental displacement field, we define the outer surface impedance matrix $\overline{\mathcal{Z} }^o\left( \lambda ,\lambda _o \right) $. This matrix complies with the ensuing pair of dimensionless differential equations:
\begin{align}\label{Eq61}
	\begin{split}
		&\frac{\mathrm{d}\overline{\mathcal{U} }\left( \lambda \right)}{\mathrm{d}\lambda}=\frac{J_g}{\lambda \left( J_g-\lambda ^3 \right)}\left( \overline{\mathcal{G} }_{11}\overline{\mathcal{U} }\left( \lambda \right) +\overline{\mathcal{G} }_{12}\overline{\mathcal{Z} }^o\left( \lambda ,\lambda _o \right) \overline{\mathcal{U} }\left( \lambda \right) \right) ,\\
		&\frac{\mathrm{d}\overline{\mathcal{Z} }^o\left( \lambda ,\lambda _o \right)}{\mathrm{d}\lambda}=\frac{J_g}{\lambda \left( J_g-\lambda ^3 \right)}\left( \overline{\mathcal{G} }_{21}+\overline{\mathcal{G} }_{22}\overline{\mathcal{Z} }^o\left( \lambda ,\lambda _o \right) -\overline{\mathcal{Z} }^o\left( \lambda ,\lambda _o \right) \overline{\mathcal{G} }_{11}-\overline{\mathcal{Z} }^o\left( \lambda ,\lambda _o \right) \overline{\mathcal{G} }_{12}\overline{\mathcal{Z} }^o\left( \lambda ,\lambda _o \right) \right) ,
	\end{split}
\end{align}
and the corresponding boundary conditions are:
\begin{align}\label{Eq62}
	\begin{split}
	\overline{\mathcal{U} }\left( \lambda _o \right) =\left[ \overline{\mathbb{U} }\left( \lambda _o \right) ,\frac{\overline{\mathbb{U} }\left( \lambda _o \right)}{\overline{\mathbb{K} }_{uv}},\left( \frac{\overline{\mathcal{Z} }_{32}^{i}\left( \lambda_o,\lambda_i \right)}{\overline{\mathcal{Z} }_{33}^{i}\left( \lambda_o,\lambda_i \right)}\frac{1}{\overline{\mathbb{K} }_{uv}}-\frac{\overline{\mathcal{Z} }_{31}^{i}\left( \lambda_o,\lambda_i \right)}{\overline{\mathcal{Z} }_{33}^{i}\left( \lambda_o,\lambda_i \right)} \right) \overline{\mathbb{U} }\left( \lambda _o \right) \right]^{\text{T}} , \quad \text{and}\quad \overline{\mathcal{Z} }^o\left( \lambda _o,\lambda _o \right) =\mathbf{0}.
	\end{split}
\end{align}
Consequently, by integrating this ensemble of differential equations (i.e., Eq. \eqref{Eq61}), subject to the initial conditions delineated in Eq. \eqref{Eq62}, we can successfully obtain the entire field of incremental displacements. It is manifest that the comprehensive Stroh-based resolution framework is exceptionally robust, effectively mitigating numerical singularities.

\begin{figure}[!htbp]
	\begin{center}
		\includegraphics[width=1\textwidth]{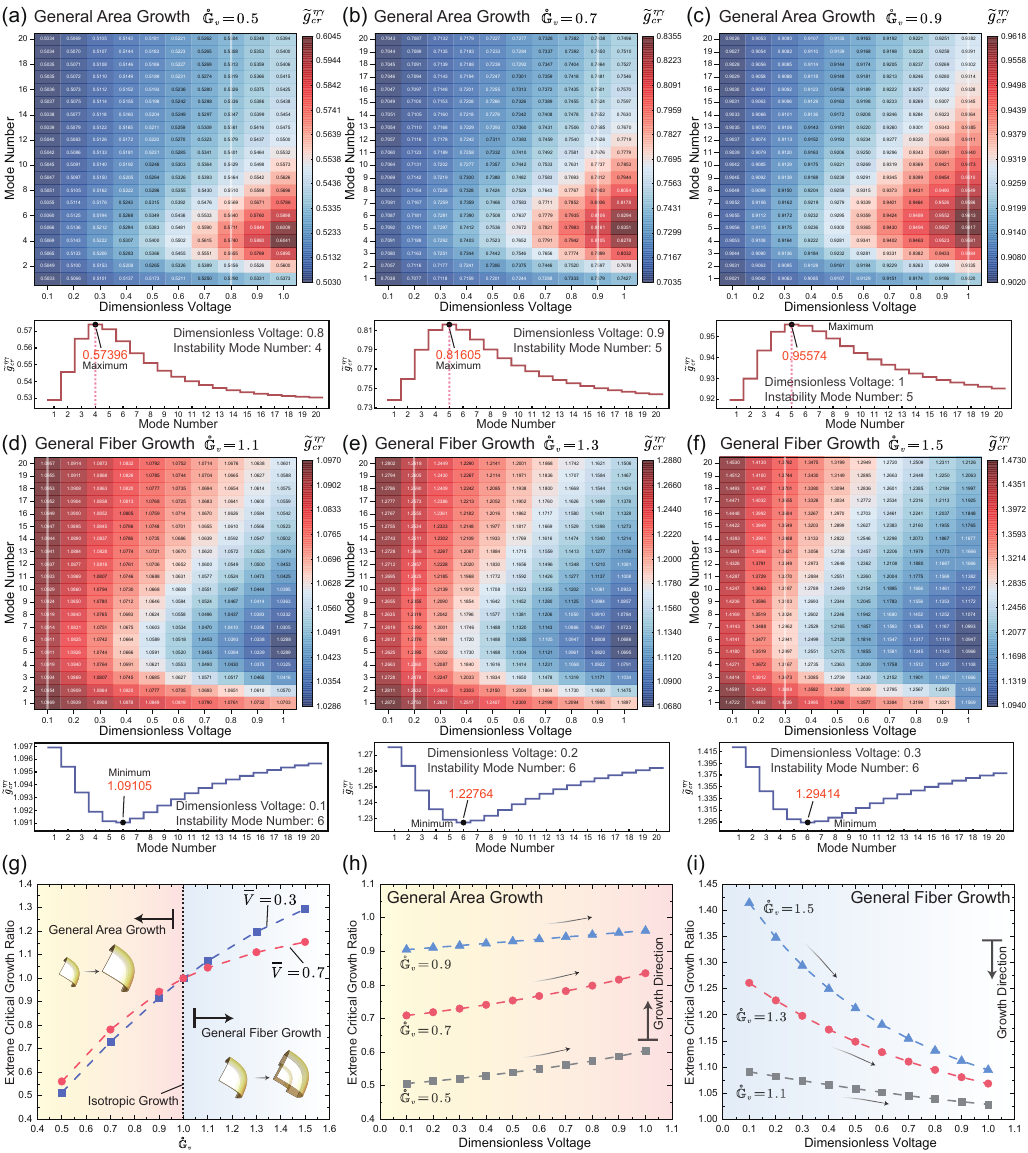}
		\caption{
			The general area growth scenario (a)-(c) demonstrates the presence of extreme Critical Growth Ratio (CGR) across various dimensionless voltages and instability mode numbers: when the growth rate ratio is
			(a) 0.5,
			(b) 0.7, and
			(c) 0.9.
			The general fiber growth scenario, spanning segments (d) to (f), highlights the prevalence of extreme CGR at diverse dimensionless voltages and instability mode numbers, with growth rate ratios of (d) 1.1, (e) 1.3, and (f) 1.5.
			(g) The nonlinear evolution of the extreme CGR with the growth rate ratio results in an image bifurcated by isotropic growth conditions: the left half showing general area growth and the right, general fiber growth.
			(h) For general area growth, and (i) for general fiber growth, the extreme CGR-dimensionless voltage curve illustrates the acceleration of biological tissue healing and regeneration through electrical stimulation.
		}\label{Fig-Perturbation-1}
	\end{center}
\end{figure}

\section{Results for electroactive healing and growth instability}\label{section5}
Utilizing the incremental field equations and the established resolution framework, we are positioned to conduct theoretical explorations into the pronounced facilitation of wound healing via electrical stimulation, the postponement of instability due to strain stiffening, as well as the morphoelastic consequences of electroactive differential growth coupled with multifactorial modulation on tissues undergoing accelerated healing. This establishes a theoretical groundwork for the development of devices engineered to expedite wound healing through electrical stimulation, anchored in empirical/experimental data.
In this section, we systematically dissect the influence of a constellation of synergistic variables on the genesis of growth-related instabilities, leading to a cascade of outcomes including the acceleration of tissue healing via electrical stimuli, the transformation in the morphologies of remodeling structures, and the emergence of delayed instabilities. 
The variables under scrutiny---differential growth dynamics across both general area growth and fiber growth, dimensionless electrostatic drivers, the coupling of mechanical and electrical fields encapsulated by the parameters \(\alpha\) and \(\beta\), internal pressurization effects, growth rate effects, and the manifestation of strain-stiffening phenomena---are integral to our understanding of the nuanced mechanisms driving biological morphogenesis and adaptive responses. 
Through our examination, we endeavor to unravel the complex interdependencies among these parameters, highlighting their collective significance in modulating the intricate dynamics of growth, stability, and morphological evolution in biological systems.

\subsection{Evolution of electroelastic differential growth}
The dynamics of tissue growth and remodeling across the vertebrate kingdom are underpinned by complex cellular differentiation processes, culminating in the emergence of structures characterized by either a unidirectional orientation or elaborate folding networks. Central to these processes are bioelectric mechanisms, with the role of endogenous electric fields and potentials being particularly critical. These intrinsic bioelectric signals, stemming from the dynamic shifts in electrochemical gradients both intra- and extracellularly, concentrate along the vectors of accelerated tissue expansion. Their interaction with growth factor activity and localized compressive stresses orchestrates cell behavior and steers growth in specific directions.
Endogenous electric fields and potentials exert a profound influence on cell proliferation, migration, and differentiation, with the magnitude and orientation of these electric cues determining cell destiny. This bioelectric guidance prompts cells to navigate towards zones of heightened electric field intensity, significantly shaping tissue morphogenesis and remodeling. Exceeding the adaptive threshold of endogenous electric fields and potentials can precipitate growth and remodeling instabilities, resulting in atypical tissue formations.
The essence of bioelectric regulation lies in its capacity not only to facilitate the differentiation of certain cell types, such as smooth muscle cells, but also to modulate the broader patterns of tissue morphological differentiation. A detailed exploration of the bioelectric phenomena at both cellular and tissue scales offers profound insights into the regulatory mechanisms at play, enhancing our comprehension of the stability, adaptability, and evolution of biological forms and functions. This deepened understanding paves the way to unlocking the underlying principles governing the intricate architectures and functionalities within biological systems. 

By integrating the notion of growth rate $\mathring{g}_i$ ($i=\gamma ,\eta $) and considering the factors outlined above, we can articulate the evolution \citep{eskandari2015systems, wang2023strain} of the growth-deformation gradient tensor as $G_{ii}\left( t \right) =1+\mathring{g}_it$ (i.e., linear representation of growth kinematics).
Hence, concerning various growth factors, we have $\gamma \left( t \right) =1+\mathring{g}_{\gamma}t$ and $\eta \left( t \right) =1+\mathring{g}_{\eta}t$. 
In order to more precisely elucidate the impact of differential growth processes, we introduce the definition of the growth rate ratio $\mathring{\mathbb{G}}_v={{\mathring{g}_{\eta}}/{\mathring{g}_{\gamma}}}$. 
If $\mathring{\mathbb{G}}_v<1$, this growth state is characterized as general area growth; conversely, if $\mathring{\mathbb{G}}_v>1$, it is delineated as general fiber growth. Conventional quantitative analyses of differential growth typically rely on a reductive simplification concerning a singular growth factor---essentially presupposing no growth in the specified direction \citep{wu2015growth}. This simplification is a matter of concern for us, as such oversimplification fails to capture the nuanced realities of differential growth phenomena within biological tissues accurately. Our proposed generalized framework for understanding differential growth enables a more nuanced and comprehensive exploration of related phenomena, including but not limited to, the delayed instability triggered by differential growth and the expedited healing processes facilitated by electrical stimulation.

Considering specific dimensionless parameters such as voltage, pressure, level of strain stiffening, and electroelastic coupling coefficients, under these conditions, if $\eta \left( t \right)>1$ is treated as a conditional parameter, it enables the expression of an alternative, distinct growth factor as $\gamma \left( t \right) =\mathring{\mathbb{G}}_{v}^{-1}\left( \eta \left( t \right) -1 \right) +1$.
Within the frameworks of general area growth ($\mathring{\mathbb{G}}_v<1$) and fiber growth ($\mathring{\mathbb{G}}_v>1$), the ordering of growth factors is distinctly defined by the conditions ${{\eta}/{\gamma}}<1$ and ${{\eta}/{\gamma}}>1$, respectively. 
Our computational strategy is centered on accurately determining the critical ratios between differential growth factors, identifying the primary bifurcation points where significant changes or transitions leading to symmetry breaking occur.
Without loss of generality, we establish the critical differential growth ratio as $\tilde{g}_{cr}^{\eta \gamma}=\left[ {{\eta}/{\gamma}} \right] _{cr}$. This definition facilitates the derivation of incremental solutions tailored to specific wrinkle wavenumbers.
Subsequently, we evaluate all possible wrinkle numbers with respect to $m$ and select the largest/smallest ratio to determine the onset of electroelastic growth instability. 
To highlight the significance of this physical quantity---that is, the maximum/minimum ratio necessary to initiate electroelastic growth instability---we introduce the concept of an extreme Critical Growth Ratio (CGR), as follows:
\begin{align}\label{Eq63}
	\begin{split}
	\left[ \tilde{g}_{cr}^{\eta \gamma} \right] _{Ext}=\begin{cases}	\max \left[ \tilde{g}_{cr}^{\eta \gamma} \right] =\max \left[ \left[ \eta /\gamma \right] _{cr} \right] \quad   &\text{(general area growth)}, \quad\\	\min \left[ \tilde{g}_{cr}^{\eta \gamma} \right] =\min \left[ \left[ \eta /\gamma \right] _{cr} \right] \,\,    &\text{(general fiber growth)} .\\\end{cases}
	\end{split}
\end{align}

Given the decoupling of the displacement field in the azimuthal direction during the solution process for incremental equations, the presence of a nonzero critical instability mode number $m_{cr}$ leads to the manifestation of three-dimensional, axisymmetric electroelastic growth instability in the system. Emphasizing the importance of this condition, the initial bifurcation of the system necessitates $m_{cr}$ to be nonzero.

\begin{figure}[!htbp]
	\begin{center}
		\includegraphics[width=1\textwidth]{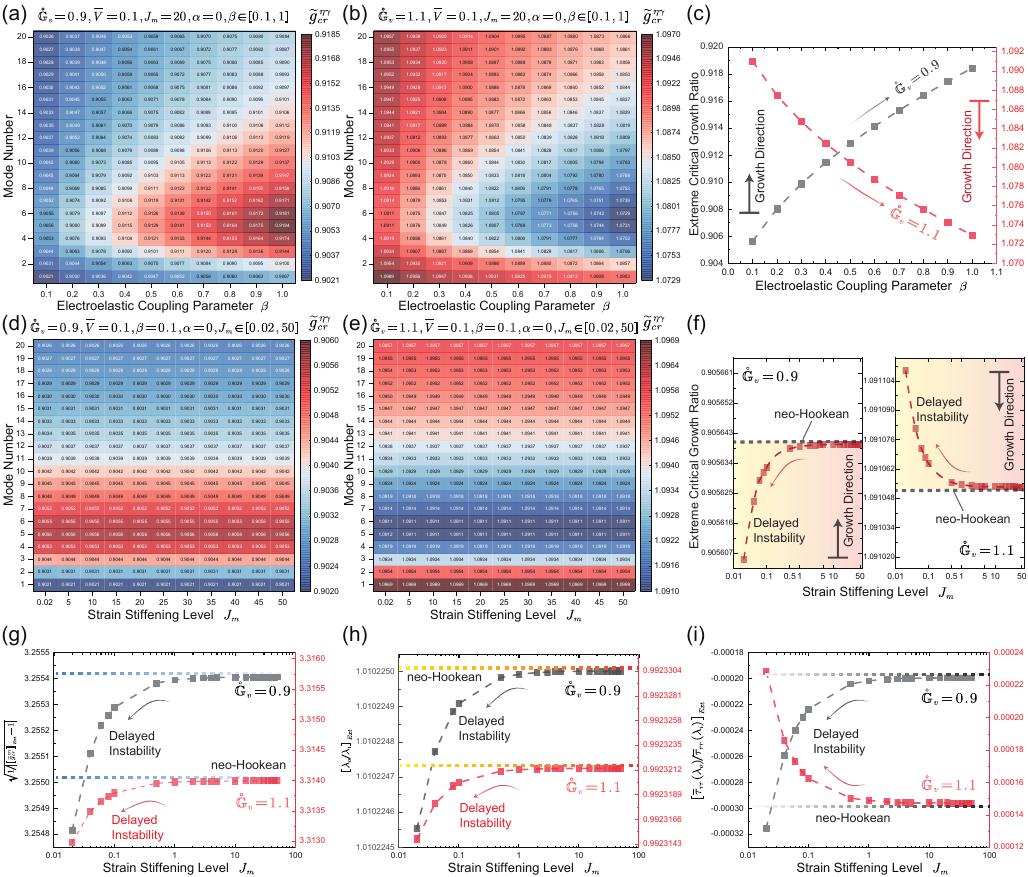}
		\caption{
			Existence of the extreme Critical Growth Ratio (CGR) for various electroelastic coupling parameters and strain stiffening levels, when the growth pattern is (a) the general area growth $\mathring{\mathbb{G}}_v=0.9$, (b) the general fiber growth $\mathring{\mathbb{G}}_v=1.1$. (c) The extreme CGR as a function of electroelastic coupling parameters. 
			The phase diagrams, delineating the correlation between the strain stiffening level and the instability mode number, provide insights into the numerical convergence of the extreme CGR.
			(f) The extreme CGR fluctuates according to the level of strain stiffening. When $J_m > 5$, the extreme CGR corresponds to the prediction of the neo-Hookean model. Conversely, for $J_m < 5$, the extreme CGR curve shifts downward for general area growth and upward for general fiber growth, resulting in a delay in growth instability compared to the neo-Hookean model.
			(g) The level of difference in extreme CGR with respect to isotropic growth conditions, $\sqrt{{{1}/{\left| \left[ \tilde{g}_{cr}^{\eta \gamma} \right] _{Ext}-1 \right|}}}$, (h) the extreme critical stretch ratio $\left[ {{\lambda_o}/{\lambda_i}} \right] _{Ext}$, and (i) the extreme critical Cauchy stress ratio $\left[ {{\bar{\tau}_{rr}\left( \lambda_o \right)}/{\bar{\tau}_{rr}}}\left( \lambda _i \right) \right] _{Ext}$ as nonlinear functions of the strain stiffening level $J_m$, reveal the delayed effect of instability in the growth of biological tissues from a deeper perspective.
		}\label{Fig-Perturbation-2}
	\end{center}
\end{figure}

\subsection{Accelerated healing and delayed instability}
In the pursuit of resolving the instability in electroelastic growth, we further segregate the overall Helmholtz free energy density into two distinct components: the elastic component $\omega$ and the electric component $\omega^*$.
Their dimensionless representation is crucial and has already been separately elaborated upon earlier. Here, in order to confine the numerical computation range of the electroelastic coupling parameters, utilizing Eq. \eqref{Eq13} as a basis, the electric field can be succinctly articulated as follows:
\begin{align}\label{Eq64}
	\begin{split}
		\mathbf{E}=\alpha \varepsilon ^{-1}\mathbf{b}_{a}^{-1}\mathbf{D}+\beta \varepsilon ^{-1}\mathbf{D}.
	\end{split}
\end{align}
After linearization of the given electric field equation and incorporating the relative dielectric permittivity, denoted as $\varepsilon_{\textit{rel}} \geq 1$, we arrive at the expression $\mathbf{E}=\varepsilon ^{-1}\varepsilon _{rel}^{-1}\mathbf{D}$. Integrating this with Eq.~\eqref{Eq64}, it follows that $\alpha (\mathbf{b}_{a}^{-1})_{11} + \beta = \varepsilon_{\textit{rel}}^{-1}$. Therefore, in scenarios where the system experiences neither deformation nor growth, the relationship simplifies to $\alpha + \beta = \varepsilon_{\textit{rel}}^{-1}$. In light of this derivation, we confine the electroelastic coupling parameters in subsequent calculations to $\alpha + \beta \leq 1$.

Fig. \ref{Fig-Perturbation-1} illustrates that for each distinct dimensionless voltage, the healing biological tissue exhibits either a maximum or minimum Critical Growth Ratio (CGR), where the maximum CGR corresponds to the scenario of general area growth, and the minimum CGR aligns with the case of general fiber growth.
During these theoretical calculations, it is essential to note that the dimensionless internal pressure ($\overline{P}=-0.001$), strain stiffening level ($J_m=20$), and electroelastic coupling parameter ($\beta=0.1$) are all finite. This observation underscores the multifactorial nature of electroelastic growth instability, wherein the interplay of these parameters plays a crucial role.
Specifically, Figs. \ref{Fig-Perturbation-1}a-c respectively demonstrate that when the growth rate ratio is 0.5, 0.7, and 0.9, the high-value region of the critical differential growth ratio $\tilde{g}_{cr}^{\eta \gamma}$ in the phase diagram composed of dimensionless voltage and instability mode number lies in the lower right quadrant. For specific voltages of 0.8, 0.9, and 1.0, as indicated by the corresponding stepped curves, it can be observed that the critical voltage and critical instability mode number at the extremes are \(m_{cr} = 4\), \(5\), and \(5\), respectively. At these points, the extreme CGR values are 0.57396, 0.81605, and 0.95574.
Correspondingly, evidence of the smallest CGR under the general fiber growth scenario, with growth rate ratios of 1.1, 1.3, and 1.5 respectively, is presented in Figs. \ref{Fig-Perturbation-1}d-f. These minimum CGRs are located in the lower right area on the dimensionless voltage-instability mode number phase diagram. For specific voltages, such as 0.1, 0.2, and 0.3, nonlinear stepped curves show that the critical instability mode number is 6 for each, with extreme CGRs of 1.09105, 1.22764, and 1.29414, respectively, indicating a progressive increase.
To circumvent additional complexities, subsequent research concentrates on elucidating the extreme CGR.
Fig. \ref{Fig-Perturbation-1}g depicts the nonlinear evolution of the extreme CGR $\left[ \tilde{g}_{cr}^{\eta \gamma} \right] _{Ext}$ with the growth rate ratio $\mathring{\mathbb{G}}_v$.
It is evident that the isotropic growth condition ($\mathring{\mathbb{G}}_v=1$, represented by the black dashed line) bifurcates the evolutionary curve into two, delineating the regions of general area growth and general fiber growth. For dimensionless voltage curves 0.3 and 0.7, higher voltages correspond to larger values in the general area growth region, while a transition occurs at the critical point ($\mathring{\mathbb{G}}_v=1$).
Expanding on this observation, as illustrated in Figs. \ref{Fig-Perturbation-1}h and \ref{Fig-Perturbation-1}i, plotting the nonlinear evolution curves of $\left[ \tilde{g}_{cr}^{\eta \gamma} \right] _{Ext}$ with dimensionless voltage in the general area growth and general fiber growth regions reveals significant effects. Particularly, under voltage stimulation, there is noticeable acceleration in the healing and growth of biological tissues.
The arrows shown in Figs. \ref{Fig-Perturbation-1}h and \ref{Fig-Perturbation-1}i signify the direction of growth and regeneration in biological tissues. General area growth and general fiber growth align with the monotonic increase and decrease, respectively, of the extreme CGR curves. Regardless of the growth mode, the trajectory of growth points (the arrows in the figure) towards isotropic conditions, suggesting that electrical stimulation expedites tissue regeneration without exhibiting preference for a specific growth mode.
The phenomena sparked by electroactive effects offer vital evidence supporting the accelerated growth, remodeling, and regeneration of biological tissues.  While we have yet to delve into molecular biology levels and have simplified the complexity of tissue properties and boundary conditions, our approach provides valuable theoretical groundwork and guidance. It sheds light on the existing anisotropic characteristics and multiscale mechanisms observed in experimental studies.

\begin{figure}[!htbp]
	\begin{center}
		\includegraphics[width=1\textwidth]{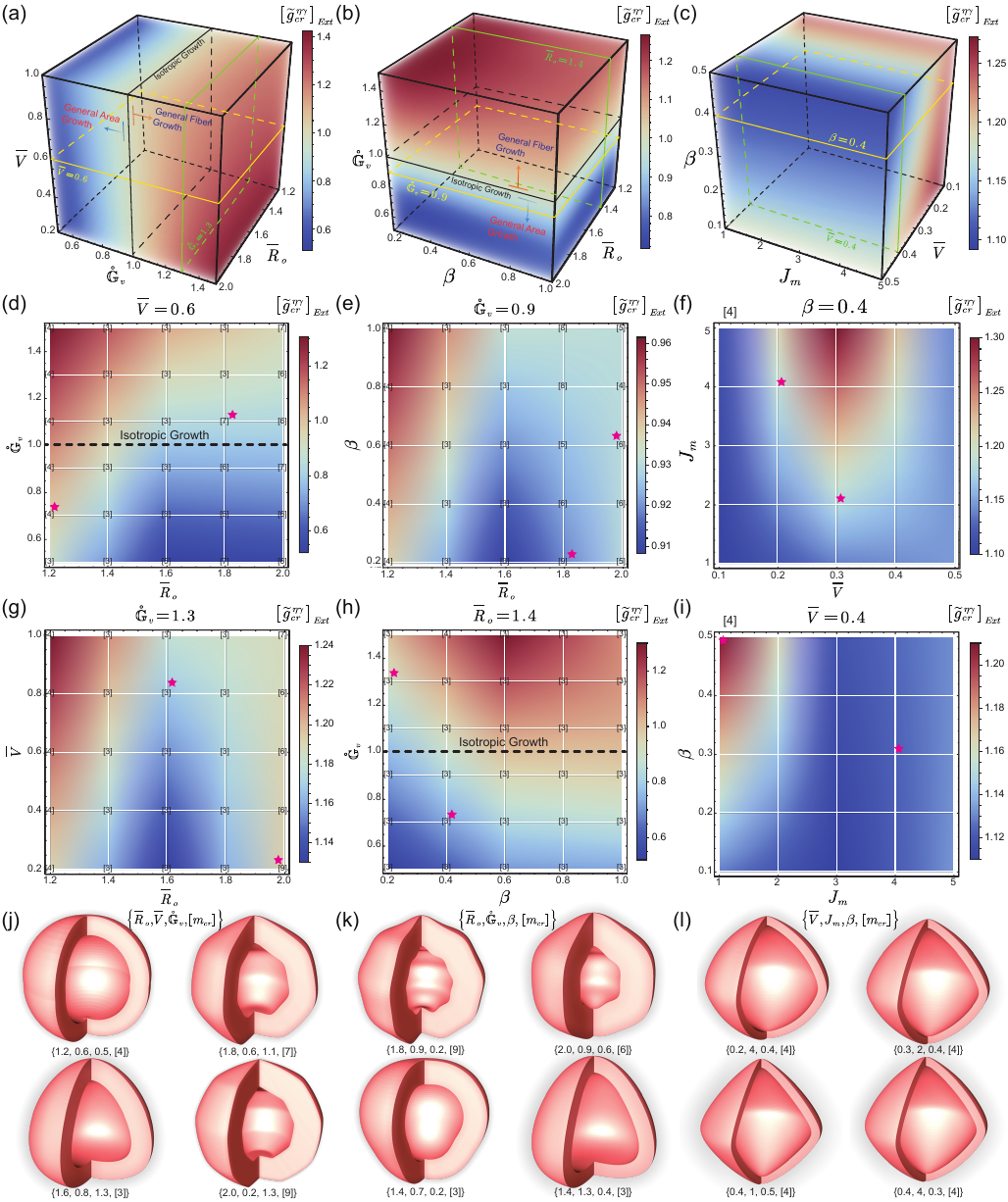}
		\caption{
			The parameter spaces influencing the extreme CGR, categorized into three sets: (a) ${\overline{R}_o,\overline{V},\mathring{\mathbb{G}}_v}$, (b) ${\overline{R}_o,\mathring{\mathbb{G}}_v,\beta}$, and (c) ${\overline{V},J_m,\beta}$. 
			Distribution maps for the extreme CGR are derived by intersecting these parameter spaces with specific planes, resulting in:
			(d) ${\overline{R}_o,\overline{V},\mathring{\mathbb{G}}_v}$ at $\overline{V}=0.6$.
			(e) ${\overline{R}_o,\mathring{\mathbb{G}}_v,\beta}$ at $\mathring{\mathbb{G}}_v=0.9$.
			(f) ${\overline{V},J_m,\beta}$ at $\beta=0.4$.
			(g) ${\overline{R}_o,\overline{V},\mathring{\mathbb{G}}_v}$ at $\beta=0.4$.
			(h) ${\overline{R}_o,\mathring{\mathbb{G}}_v,\beta}$ at $\overline{R}_o=1.4$.
			(i) ${\overline{V},J_m,\beta}$ at $\overline{V}=0.4$.
			Electroelastic instability modes, marked by red stars on the maps, are summarized:
			(j) Modes for (d) and (g),
			(k) modes for (e) and (h),
			(l) modes for (f) and (i).
		}\label{Fig-Perturbation-3}
	\end{center}
\end{figure}

The parameter $\beta$ quantifies the coupling strength between mechanical forces and electrical fields, with a higher $\beta$ reflecting enhanced performance in device applications due to stronger electroelastic interactions. The parameter $J_m$ is a measure of strain stiffening within the material model, where a decrease in $J_m$ leads to a pronounced steepening of the stress-strain response under minimal applied strain, signifying a more "rigid" material behavior. Conversely, a larger $J_m$ suggests a softer material characteristic. As $J_m$ approaches positive infinity, the material behavior simplifies to that of the neo-Hookean model, highlighting the loss of strain stiffening effects. 

Fig.~\ref{Fig-Perturbation-2} illustrates the presence of extreme Critical Growth Ratios (CGR) across different electroelastic coupling strengths $\beta$ and strain stiffening levels $J_m$, showcasing the phenomena of delayed instability triggered by strain stiffening.
Specifically, for $\overline{V}=0.1$, $J_m=20$, and $\alpha =0$, Figs.~\ref{Fig-Perturbation-2}a-c illustrate the nonlinear evolution patterns of the extreme CGR under specific electroelastic coupling parameters $\beta \in \left[ 0.1,1 \right] $ and instability mode numbers $m$. Figs.~\ref{Fig-Perturbation-2}a and \ref{Fig-Perturbation-2}b correspond to the general area growth and the general fiber growth scenarios, respectively, with growth rate ratios of 0.9 and 1.1. It is observed that the maximum and minimum values of the CGR generally occur in the lower right corner of the corresponding phase diagrams. When specifying certain electroelastic coupling parameters, the nonlinear stepped curves of the extreme CGR exhibit convex or concave characteristics.
Fig.~\ref{Fig-Perturbation-2}c illustrates the nonlinear evolution pattern of the extreme CGR concerning electroelastic coupling parameters, showcasing its impact on both general area growth and general fiber growth scenarios. Larger values of $\beta$ contribute to accelerated tissue growth, aligning the growth direction with isotropic growth conditions. This proactive promotion resembles the growth acceleration induced by voltage stimulation but carries deeper theoretical implications:
Greater electroelastic coupling parameters typically correlate with superior device performance, facilitating a more efficient conversion between electrical and mechanical energy interactions. 
As the electroelastic coupling parameter $\beta$ increases, the phenomenon of tissue contraction and strengthening along the direction of the electric field becomes more pronounced. 
This growth and reinforcement not only improve the mechanical integrity of the tissue but also encourage cellular migration and proliferation. Further increases in the electroelastic coupling parameter $\beta$ accelerate tissue growth in alignment with the electric field, thereby streamlining the overall growth process of biological tissues.
Subsequently, Figs.~\ref{Fig-Perturbation-2}d and \ref{Fig-Perturbation-2}e illustrate the interplay between the extreme CGR and the level of strain stiffening under scenarios of general area growth ($\mathring{\mathbb{G}}_v=0.9$) and general fiber growth ($\mathring{\mathbb{G}}_v=1.1$), respectively. Notably, across the corresponding phase diagrams, the numerical values of the extreme CGR remain relatively constant in the horizontal direction, suggesting a convergence property of the extreme CGR.
To delve deeper into the observed convergence, the analysis presented in Fig.~\ref{Fig-Perturbation-2}f sheds light on the dynamics between the extreme CGR and varying levels of strain stiffening. Specifically, it demonstrates that, for strain stiffening levels exceeding $J_m > 5$, the growth trajectories for both the general area and fiber scenarios align with the curve indicative of a neo-Hookean material model. This alignment signifies a convergence of growth patterns under high strain stiffening levels. In contrast, a pronounced deflection phenomenon becomes evident as $J_m$ is reduced from 5 to 0.02. This is characterized by a downward deflection in the curve for general area growth and an upward deflection for general fiber growth, emphasizing the collective influence exerted on the extreme CGR by levels of strain stiffening---particularly at lower values of $J_m$---and differential growth effects.
From a mechanics viewpoint, the significant deviation of these curves indicates an important phenomenon: the delay in instability triggered by strain stiffening. Imagining a scenario in which biological tissues lack strain stiffening effects, the extreme CGR can be uniquely defined, and its theoretical value can be derived using the neo-Hookean model. However, with noticeable strain stiffening present, biological tissues gain reinforcement at lower applied strains. This leads to a postponement in the emergence of tissue growth instability compared to that predicted by neo-Hookean materials, a scenario vividly captured by the substantial deflection of the extreme CGR curves. Thus, conventional research methods that disregard the impact of strain stiffening often result in significant discrepancies, either overestimating or underestimating, in the quantitative analysis of growth instability in biological tissues.
To illuminate the delayed instability phenomenon that strain stiffening prompts, we conduct a detailed examination of the dynamics involving geometry, stress, and differential growth modalities, as delineated in Figs.~\ref{Fig-Perturbation-2}g-i. Specifically, Fig.~\ref{Fig-Perturbation-2}g delineates the departure of the extreme CGR $\sqrt{{{1}/{| [ \tilde{g}_{cr}^{\eta \gamma} ] _{Ext}-1 |}}}$ from isotropic growth benchmarks in scenarios of general area and fiber growth. Notably, when $J_m$ exceeds 5, the trajectories align with those predicted by the neo-Hookean model, indicating convergence. Conversely, for $J_m$ below 5, we observe a pronounced postponement in the initiation of instability due to differential growth. 
This trend mirrors in Fig.~\ref{Fig-Perturbation-2}h, showcasing a similar pattern in the peak of the critical stretch ratio $\left[ {{\lambda _o}/{\lambda _i}} \right] _{Ext}$, signifying a lag in growth instability for $J_m$ values below this threshold. Furthermore, Figure~\ref{Fig-Perturbation-2}i elucidates the extreme critical Cauchy stress ratio for the growth of biological tissue, represented by $\left[ {{\bar{\tau}_{rr}\left( \lambda _o \right)}/{\bar{\tau}_{rr}}}\left( \lambda _i \right) \right] _{Ext}$. As the level of strain stiffening increases, the material properties become more compliant, leading to a convergence of  $\left[ {{\bar{\tau}_{rr}\left( \lambda _o \right)}/{\bar{\tau}_{rr}}}\left( \lambda _i \right) \right] _{Ext}$ values, ultimately aligning with those of neo-Hookean materials.

Expanding upon this groundwork, the accumulation of collagen at wound sites during the remodeling phase leads to strain stiffening, shedding light on the underlying mechanisms at play. As the level of strain stiffening increases within the tissue, making the tissue softer, strain stiffening more effectively promotes tissue growth. In contrast, a lower level of strain stiffening, indicative of harder tissue, results in a marked delay in the instability of wound tissue. This postponed instability helps to diminish surface wrinkling at the wound site, resulting in a smoother wound surface that facilitates wound closure. This interplay between strain stiffening and tissue response not only highlights the mechanics aspects of tissue growth and wound healing but also points to potential strategies for managing and optimizing healing processes.

\begin{figure}[!t]
	\begin{center}
		\includegraphics[width=1\textwidth]{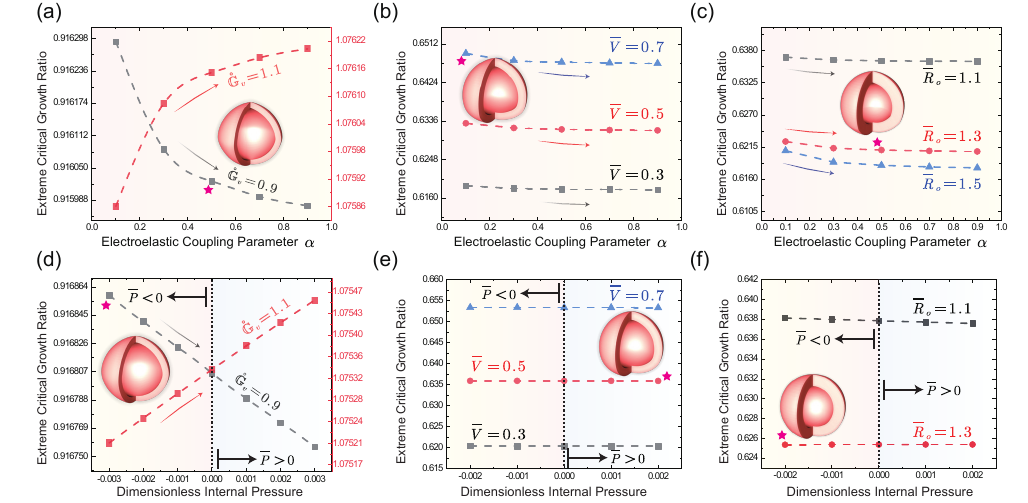}
		\caption{
			In the context of variable parameters (a) $\mathring{\mathbb{G}}_v$, (b) $\overline{V}$, and (c) $\overline{R}_o$, the extreme CGR manifests nonlinear evolution patterns as a function of the electroelastic coupling coefficient $\alpha$. 
			Likewise, under the conditions of (d) $\mathring{\mathbb{G}}_v$, (e) $\overline{V}$, and (f)  $\overline{R}_o$, the critical CGR exhibits a distinct nonlinear evolutionary response to dimensionless internal pressure $\overline{P}$.
		}\label{Fig-Perturbation-4}
	\end{center}
\end{figure}

Now, our investigations have elucidated the role of electrical stimulation in augmenting the proliferation and facilitating the mitigation of delayed instability phenomena precipitated by strain stiffening within the context of biological tissue growth, healing, and regeneration. Nonetheless, a paramount challenge persists in delineating the mechanisms through which strain stiffening magnitude and nondimensional voltage parameters critically modulate tissue growth dynamics and morphogenetic pathways.
Under the constraint that $\alpha + \beta \leq 1$ (wherein $\alpha$ is initially set to $0$, a limitation to be subsequently alleviated), and confined to the regime where strain stiffening precipitates delayed instability ($J_m \in [0.02, 5]$), we have rigorously defined the parameter space delineating the mechanisms of growth, healing, and instability phenomena in biological tissues. This comprehensive mapping is illustrated in Fig.~\ref{Fig-Perturbation-3}.
More specifically,in the depicted extreme CGR parameter space of Fig.~\ref{Fig-Perturbation-3}a, encompassing variables $\overline{V}$, $\mathring{\mathbb{G}}_v$, and the radius ratio $\overline{R}_o$, and bifurcated by the isotropic growth condition (i.e., $\mathring{\mathbb{G}}_v=1$) into sectors representing general area ($\mathring{\mathbb{G}}_v<1$) and fiber growth ($\mathring{\mathbb{G}}_v>1$), we initiate our analysis by intersecting this space with a predefined $\overline{V}$ plane ($\overline{V}=0.6$), yielding Fig.~\ref{Fig-Perturbation-3}d. 
This intersection reveals a coupling effect between $\mathring{\mathbb{G}}_v$ and the radius ratio $\overline{R}_o$, whereby larger and smaller extreme CGR values are discernibly positioned in the upper left and lower right quadrants of Fig.~\ref{Fig-Perturbation-3}d, respectively. 
Advancing further, the parameter space of Fig.~\ref{Fig-Perturbation-3}a is bisected by a vertical plane at $\mathring{\mathbb{G}}_v=1.3$, culminating in Fig.~\ref{Fig-Perturbation-3}g. 
This section clearly delineates that the upper left corner harbors larger extreme CGR values, whereas the domain characterized by $\overline{R}_o=1.6$ and $\overline{V}=1.2$ houses smaller extreme CGR values. 
To delve deeper into morphological electroelasticity, Fig.~\ref{Fig-Perturbation-3}j elucidates the selection of instability patterns in electroelastic growth, highlighted by red stars in Figs.~\ref{Fig-Perturbation-3}d and \ref{Fig-Perturbation-3}g, with the pertinent parameters specified by the set $\{ \overline{R}_o,\overline{V},\mathring{\mathbb{G}}_v,[ m_{cr} ] \} $.
In Fig.~\ref{Fig-Perturbation-3}b, the parameter space delineating extreme CGR is constructed utilizing variables $\mathring{\mathbb{G}}_v$, $\beta$, and $\overline{R}_o$. A plane defined at $\mathring{\mathbb{G}}_v=1$ demarcates two distinct differential growth modes, enhancing our understanding of the growth dynamics. To further elucidate the coupling relationships between these pivotal variables, the parameter space is initially intersected with a plane at $\mathring{\mathbb{G}}_v=0.9$, yielding Fig.~\ref{Fig-Perturbation-3}e. This intersection reveals that higher extreme CGR values are predominantly located in the upper left quadrant of the $\beta$-$\overline{R}_o$ plane, whereas lower extreme CGR values cluster in the central to lower sections of the figure.
Further exploration involves intersecting the parameter space from Fig.~\ref{Fig-Perturbation-3}b with a plane at $\overline{R}_o=1.4$, resulting in Fig.~\ref{Fig-Perturbation-3}h. Here, higher extreme CGR values occupy the upper middle region, and lower values are found in the lower left corner, underscoring the complex nonlinear interplay among these critical variables.
In a manner akin to the analysis presented in Fig.~\ref{Fig-Perturbation-3}j, we theoretically delineate the first bifurcation forms of electroelastic growth instability (refer to $\{\overline{R}_o,\mathring{\mathbb{G}_v},\beta,[m_{cr}]\}$ in Fig.~\ref{Fig-Perturbation-3}k for details), as indicated by red stars in Figs.~\ref{Fig-Perturbation-3}e and ~\ref{Fig-Perturbation-3}h.
Next, in Fig.~\ref{Fig-Perturbation-3}c, we investigate the synergistic effects of $\beta$, $J_m$, and $\overline{V}$ on the extreme CGR.
A specific cross-section at $\beta=0.4$ isolates the interaction between the strain stiffening level $J_m$ and the dimensionless voltage $\overline{V}$, as illustrated in Fig.~\ref{Fig-Perturbation-3}f. This section reveals a region slightly above the center with elevated extreme CGR values, contrasting with the diminished values in the lower left corner. Importantly, all observed values exceed 1, signifying that each state within Fig.~\ref{Fig-Perturbation-3}f corresponds to general fiber growth. The states of electroelastic growth instability, denoted as $\{ \overline{V},J_m,\beta ,[ m_{cr} ] \}$ and marked by red stars, are detailed in the initial segments of Fig.~\ref{Fig-Perturbation-3}i.
Further analysis entailed slicing the parameter space of Fig.~\ref{Fig-Perturbation-3}c at a dimensionless voltage of 0.4 to derive Fig.~\ref{Fig-Perturbation-3}i, highlighting the nonlinear interaction between $\beta$ and $J_m$. The upper left quadrant of this figure contains higher extreme CGR values, indicating a susceptibility to electroelastic growth instability. The instability states $\{\overline{V},J_m,\beta ,[ m_{cr} ]\}$ identified with red stars are depicted within Fig.~\ref{Fig-Perturbation-3}i.

The principal physical significance of phenomena such as electroelastic growth instability, the acceleration of tissue growth through electrical stimulation, and the delayed instability induced by strain stiffening lies in their intricate regulation by multiple factors. 
This complexity is systematically unraveled in Fig.~\ref{Fig-Perturbation-3}, where we examine the collaborative impact of $\mathring{\mathbb{G}}_v$, $J_m$, $\beta$, and $\overline{V}$ on the extreme CGR. 
It is crucial to underscore that in this exploration, we initially set the electroelastic coupling parameter, $\alpha$, to zero and maintained a consistent dimensionless internal pressure across the analysis. 
Moving beyond these initial parameters (see Fig.~\ref{Fig-Perturbation-4}), our investigation pivots to scrutinizing the effects of varying the electroelastic coupling parameter, $\alpha$, alongside the dimensionless internal pressure $\overline{P}$, thereby shedding light on their roles in modulating the electroelastic growth and associated instabilities in biological tissues.
In Fig.~\ref{Fig-Perturbation-4}a, the nonlinear trajectory of the extreme CGR as a function of the electroelastic coupling parameter $\alpha$ is depicted, under conditions specified by $\beta=0.1$, $J_m=20$, $\overline{R}_o=1.2$, $\overline{V}=0.4$, and $\overline{P}=-0.001$. 
The depicted black and red lines correspond to the patterns of general area growth ($\mathring{\mathbb{G}}_v=0.9$) and general fiber growth ($\mathring{\mathbb{G}}_v=1.1$), respectively. 
Notably, an increase in the electroelastic coupling parameter a appears to suppress tissue growth, a phenomenon that stands in stark contrast to the influence exerted by $\beta$. 
This finding prompts a further examination, carried out in Fig.~\ref{Fig-Perturbation-4}b and \ref{Fig-Perturbation-4}c, where $\mathring{\mathbb{G}}_v$ is held constant while $\overline{V}$ is varied to probe deeper into the ramifications of a on the extreme CGR. 
It emerges that for given values of the electroelastic coupling parameter $\alpha$, an escalation in voltage and a reduction in the radius ratio lead to an enhancement of the extreme CGR. 
While, this relationship exhibits a nuanced complexity; specifically, for fixed values of $\overline{V}$ and $\overline{R}_o$, the extreme CGR demonstrates a gradual decline, suggesting a diminished sensitivity to higher values of $\alpha$.
The narrative progresses as we explore the influence of dimensionless internal pressure $\overline{P}$ on the extreme CGR in Figs.~\ref{Fig-Perturbation-4}d-f. 
Fig.~\ref{Fig-Perturbation-4}d reveals that an evolution from negative to positive dimensionless internal pressure fosters electroelastic growth. 
Intriguingly, across a particular dimensionless voltage range, escalating internal pressure does not substantially alter the extreme CGR, pointing to a relative insensitivity. 
This trend of insensitivity persists even when considering a fixed radius ratio, as delineated in Fig.~\ref{Fig-Perturbation-4}f, further complicating the landscape of electroelastic growth and instability in biological tissues.
In Summary, this investigation thoroughly examines the effects of electroelastic activation, the heterogeneity of growth patterns, and the phenomenon of strain stiffening on tissue growth, alongside the processes of electroelastic remodeling and regeneration. The study employs a rigorously supported model to illuminate the processes that facilitate accelerated healing in response to electrical stimulation, as well as the onset of delayed instability within electroelastic growth, underscoring their profound implications for clinical practices. 
Despite the reductionist approach towards the complexity of tissue characteristics and boundary conditions, it lays down a critical theoretical groundwork for the advancement of research into more sophisticated models and the exploration of multiscale mechanisms. 
The significance of this research lies in its contribution to the scientific understanding of the experimental mechanisms at play.

\section{Discussion and conclusions}\label{section6}
Motivated by experimental insights into differential growth and electroelastic remodeling facilitating rapid healing in biological tissues, we introduce a sophisticated theoretical model elucidating electroactive differential growth, mass reconfiguration, and morphological instability. 
Demonstrative investigations are pursued through experiments on electroactive differential growth, juxtaposing spontaneous wound recovery with the expedited healing propelled by external electrical stimuli, underpinning clinical incentives.
Our theoretical discourse decomposes biological growth into multiplicative components of the total deformation gradient, commencing from a pristine stress-free baseline and integrating electroelastic and strain stiffening into the growth deformation framework. This model comprehensively details the Helmholtz free energy density and electroelastically coupled Cauchy stress in growing tissues, incorporating general area and fiber growth alongside nonlinear strain stiffening phenomena. It facilitates a methodical exploration of differential growth patterns and the breaking of symmetry in biological tissues, enhanced by electrical stimulation for swift healing.
Employing perturbation theory, we explore the evolutionary dynamics of growth, theoretically substantiating the beneficial influence of electroactivity on tissue healing. 
The study reveals how electric and elastic fields interact synergistically, influencing differential growth and strain stiffening to mold tissue morphology, which significantly delays the onset of electroelastic instability.
Despite the pivotal role of spatiotemporal oscillatory patterns in tissue healing, our application of nonlinear field theory unveils the nuanced dynamics of differential growth and the subsequent delay in initial morphological bifurcation triggered by electroactivity and strain stiffening, presenting profound implications for scientific understanding.

Initially, we address the imperative of the Kröner-Lee decomposition, which delineates material deformation into distinct elastic and growth factions. Through the application of Stokes' theorem and Gauss's divergence theorem, we elucidate the interchangeability of electric field intensity and electric displacement vectors in diverse configurations. This methodology fosters the formulation of Cauchy stress and total nominal stress equations, culminating in the identification of a reduced form for the energy function.

Secondly, we conceptualize the electroactive tissue wound as an evolving thick spherical shell structure, characterized by a voltage differential and internal pressure. 
Leveraging the isochoric principle, we ascertain a streamlined version of the first Cauchy equation. 
By segregating the total Helmholtz free energy density into its elastic and electroelastic segments, we formulate expressions for dimensionless voltage and electric displacement. 
Following this, we employ the Gent model, distinguishing growth into three distinct categories. The ground state outcomes are meticulously analyzed, incorporating considerations of differential growth, dimensionless voltage, and the impact of strain stiffening.

Subsequently, employing perturbation theory, we delve into the electroelastic remodeling and morphological bifurcation behavior of spherical shell structures endowed with electroactivity and strain stiffening effects.
By initiating with first-order incremental kinematic relations, we succeed in deriving superimposed incremental equilibrium equations alongside incremental Maxwell's equations. This enables the transformation of the electroelastic growth bifurcation problem into a Stroh resolution framework via the formulation of an incremental displacement field. Following this, the surface impedance method is applied to meticulously evaluate the comprehensive incremental displacement and incremental traction fields.

In the analysis of electroelastic growth instability, our study elucidates three pivotal aspects of tissue growth under electrical stimulation:
(1) Voltage stimulation significantly accelerates the healing and growth of biological tissues. This phenomenon suggests that tissue regeneration is expedited across various growth modes, encouraging a transition towards isotropic conditions without showing a preference for specific growth pathways. The accelerated growth and remodeling, facilitated by electroactive effects, underscore the potential of electrical stimulation in enhancing tissue regeneration.
(2) The performance of devices used in electrical stimulation is enhanced with larger electroelastic coupling parameters. A greater electroelastic coupling parameter induces contraction and reinforcement of biological tissues along the direction of the electric field. This contraction and reinforcement promote cell migration and proliferation, with larger electroelastic coupling parameters speeding up tissue growth in the direction of growth.
(3) During the remodeling phase, collagen accumulation at wound sites introduces strain stiffening, which subtly influences tissue growth. An increased level of strain stiffening softens the tissue and promotes growth, while lower levels of strain stiffening result in harder tissue, leading to a significant delay in tissue instability at the wound site. This delayed instability helps reduce surface wrinkling at the wound site, resulting in a smoother wound surface conducive to wound closure.

Finally, it is imperative to acknowledge that while tissue healing for many is a trivial concern, resolving with minimal intervention, for individuals suffering from chronic conditions like diabetes mellitus, peripheral vascular disease, or compromised immune systems (e.g., systemic lupus erythematosus), and those impacted by poor nutrition or aging, acute wounds have a propensity to become chronic \citep{kapp2017financial, shirzaei2023stretchable}. These chronic wounds represent a significant challenge in tissue regeneration, contributing to a substantial socioeconomic burden \citep{grzelak2023pharmacological}. They account for an estimated 1\% to 3\% of total healthcare expenditures in developed countries, a figure that is increasing in tandem with the median age of the population \citep{olsson2019humanistic}. Moreover, chronic wounds impose severe discomfort and limitations on those affected.
Our exploratory journey begins with the experimental application of conductive polymer and 3D-printed bioelectric devices aimed at enhancing tissue healing rates. This initial phase leads to the development of theories centered on accelerated tissue regeneration, grounded in nonlinear electroelastic theory and nonlinear field theory \citep{dorfmann2014nonlinear}. Our work illuminates phenomena such as expedited growth and variable growth rates. While our theoretical and computational explorations do not delve into subcellular mechanisms, the framework we have developed provides crucial insights into the mechanisms driving electroactive tissue growth and morphological instabilities.


\section*{Declaration of Competing Interest}
The authors declare that they have no known competing financial interests or personal relationships that could have appeared to influence the work reported in this paper.

\section*{Acknowledgments}
This research received funding from the National Natural Science Foundation of China under Grant Nos. 12202105, 12122204, 12372096, 12172102, and 52373139. 
This work is also partly supported by the Basic Research Program of Shenzhen (Grant No. 20231116101626002), Scientific Research Platforms and Projects funded by the Guangdong Provincial Education Office (Grant No. 2022ZDZX3019), Shanghai Pilot Program for Basic Research at Fudan University (Grant Nos. 21TQ1400100-21TQ010), Shanghai Shuguang Program (Grant No. 21SG05), and the Guangdong Basic and Applied Basic Research Foundation (Grant No. 2023A1515110532). Additional support was provided by the specialized research projects of the Huanjiang Laboratory in Zhuji, Zhejiang Province.
Y.D. also acknowledges support from the European Union GA n$^\circ$101105740 "Multi-scale and Multi-physics Modelling of Soft Tissues - MULTI-SOFT". (The views and opinions expressed are those of the author(s) only and do not necessarily reflect those of the European Union or the European Research Executive Agency. Neither the European Union nor the granting authority can be held responsible for them.)

\appendix
\section*{Appendix A. Bioelectric Device and Wound Morphological Evolution}\label{AppendixA}
Experimental Methods: All relevant materials are sterilized, including portable batteries, bioelectric stimulation devices, wound dressings, and others. Female SD rats weighing 150-200 grams are used to evaluate the effect of the bioelectric stimulation device and portable batteries on in vivo skin wound healing. Two symmetrical smooth areas on the dorsal side of the rats are selected to create wounds, allowing for the fixation of the bioelectric stimulation device and portable batteries to generate an electric field around the wound. Prior to surgery, the dorsal area of each rat is completely depilated, and a hair removal cream is applied to keep the wound site clean. After anesthesia, an 8 mm diameter circular full-thickness skin layer is removed from the back of the rats using a skin biopsy punch.
Two circular wounds are created on the smooth dorsal area of the experimental rats: the left side serves as the control group (without any treatment) and the right side as the bioelectric stimulation group. In the control group, wound dressings are fixed at the wound site using medical tape. In the bioelectric stimulation group, the electrode consists of a circular conductive silver gel (positive electrode) and a central point electrode (negative electrode), connected to the positive and negative terminals of the portable battery, respectively. The electrode is fixed near the edge of the wound (perpendicular to the rat body axis) with medical tape to prevent the device from falling off. The electrical stimulation protocol involves stimulating once a day for one hour (1V). After each electrical stimulation session, the device is removed, and a wound dressing is applied to protect the wound from inflammation and suppuration.
Wound healing in the experimental and control groups is recorded by taking photographs of the wounds on days 0, 1, 2, 3, 4, 5, and 6. The degree of wound healing is measured based on the preserved size of the wound, considering slight variations in the initial wound size. The degree of wound closure is evaluated based on its relative size to that on day 0. After the experiment, tissue from the electrode site is collected for histological analysis to characterize its wound healing repair effect. The experimental procedures and observation times for the control group without the electric field are the same as those for the experimental group. Wound areas are measured in photographs using ImageJ software.
All animal surgeries are approved by the Committee on Animal Care at SUSTech, Protocol No. SUSTech-JY202312002. All experimental personnel must undergo professional training, wear protective clothing, and be fully disinfected.

\renewcommand{\thefigure}{A\arabic{figure}}
\setcounter{figure}{0}
\begin{figure}[!h]
	\begin{center}
		\includegraphics[width=1\textwidth]{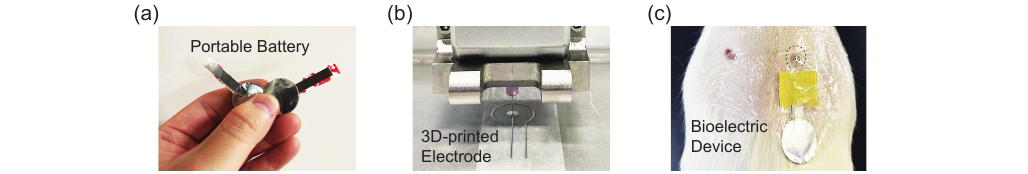}
		\caption{
			(a) A portable battery capable of providing a 1V voltage.
			(b) Utilization of 3D printing technology and conductive polymers to fabricate electrode.
			(c) Assembly of the portable battery with the 3D-printed electrode and formation of a bioelectric device for treating wounds in SD rats (ES group), with the wound on the left side serving as the Control group.
		}\label{Fig-A1}
	\end{center}
\end{figure}

\begin{figure}[!h]
	\begin{center}
		\includegraphics[width=1\textwidth]{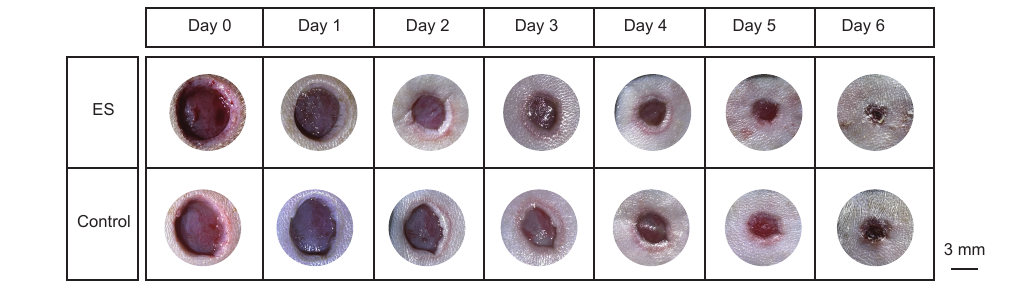}
		\caption{
		 Representative growth and morphological instability evolution of wounds in the ES and Control groups of SD rats.
		 Evidently, wounds in the ES group heal more rapidly, demonstrating a higher healing rate and a smaller wound area by day six.
		}\label{Fig-A2}
	\end{center}
\end{figure}

\section*{Appendix B. Sub-matrix Configuration of the Dimensionless Stroh Matrix}\label{AppendixB}
\begin{align}\label{Eq A1}
	\begin{split}
		&\overline{\mathcal{G} }_{11}=\left[ \begin{matrix}	-2&		\kappa&		0\\	{{\left( \overline{\tau }_{rr}-\overline{\omega } \right) \kappa} /{\overline{\omega }}}&		{{\left( \overline{\omega }-\overline{\tau }_{rr} \right)} /{\overline{\omega }}}&		0\\	-{{\overline{\tau }_{rr}\overline{\varrho }\kappa ^2} /{\overline{\omega }}}&		{{\overline{\tau }_{rr}\overline{\varrho }\kappa} /{\overline{\omega }}}&		-1\\\end{matrix} \right] , \quad \overline{\mathcal{G} }_{12}=\left[ \begin{matrix}	0&		0&		0\\	0&		{{1} /{\overline{\omega }}}&		{{\overline{\varrho }\kappa} /{\overline{\omega }}}\\	0&		-{{\overline{\varrho }\kappa} /{\overline{\omega }}}&		\left( -{{\overline{\varrho }^2} /{\overline{\omega }}}-{{1} /{\overline{\mathcal{K} }_{022}^{*}}} \right) \kappa ^2\\\end{matrix} \right] ,\\&\overline{\mathcal{G} }_{21}=\left[ \begin{matrix}	\left( \overline{\varsigma }-{{\left( \overline{\tau }_{rr}-\overline{\omega } \right) ^2} /{\overline{\omega }}} \right) \kappa ^2+2\overline{\varLambda }&		\left( {{\left( \overline{\omega }-\overline{\tau }_{rr} \right) ^2} /{\overline{\omega }}}-\overline{\varsigma }-\overline{\varLambda } \right) \kappa&		2\overline{\mathcal{T} }\\	\left( {{\left( \overline{\omega }-\overline{\tau }_{rr} \right) ^2} /{\overline{\omega }}}-\overline{\varsigma }-\overline{\varLambda } \right) \kappa&		-\left( \kappa ^2\overline{\varDelta }+\overline{\varUpsilon }-\overline{\varsigma }+{{\left( \overline{\omega }-\overline{\tau }_{rr} \right) ^2} /{\overline{\omega }}} \right)&		-\kappa \overline{\mathcal{T} }\\	-2\overline{\mathcal{T} }&		\kappa \overline{\mathcal{T} }&		-\overline{\mathcal{K} }_{011}^{*}\\\end{matrix} \right] ,  \\&\overline{\mathcal{G} }_{22}=\left[ \begin{matrix}	1&		{{\left( \overline{\omega }-\overline{\tau }_{rr} \right) \kappa} /{\overline{\omega }}}&		-{{\overline{\tau }_{rr}\overline{\varrho }\kappa ^2} /{\overline{\omega }}}\\	-\kappa&		{{\left( \overline{\tau }_{rr}-2\overline{\omega } \right)} /{\overline{\omega }}}&		{{\overline{\tau }_{rr}\kappa \overline{\varrho }}/{\overline{\omega }}}\\	0&		0&		0\\\end{matrix} \right],
	\end{split}\tag{A1}
\end{align}
with
\begin{align}\label{Eq A2}
	\begin{split}
		&\kappa =\sqrt{m\left( m+1 \right)}, \quad \overline{\omega }=\left( \overline{\mathcal{A} }_{01212}^{*}-\frac{\overline{\varGamma }_{0122}^{*2}}{\overline{\mathcal{K} }_{022}^{*}} \right) , \quad \overline{\varrho }=\left( \frac{\overline{\varGamma }_{0122}^{*}}{\overline{\mathcal{K} }_{022}^{*}} \right) , \quad \overline{\varsigma }=\left( \overline{\mathcal{A} }_{02121}^{*}-\frac{\overline{\varGamma }_{0122}^{*2}}{\overline{\mathcal{K} }_{022}^{*}} \right) ,\\&\overline{\varLambda }=\left( 2\overline{\mathcal{A} }_{01111}^{*}-4\overline{\mathcal{A} }_{01122}^{*}+\overline{\mathcal{A} }_{02222}^{*}+\overline{\mathcal{A} }_{02233}^{*}+3\overline{q} \right) , \quad \overline{\varDelta }=\left( 2\overline{\mathcal{A} }_{01122}^{*}-\overline{\mathcal{A} }_{01111}^{*}-\overline{\mathcal{A} }_{02222}^{*}-2\overline{q} \right),  \\&\overline{\varUpsilon }=\left( \overline{\mathcal{A} }_{02222}^{*}-\overline{\mathcal{A} }_{02233}^{*}+\overline{q} \right), \quad \overline{q}=\left(\overline{\mathcal{A} }_{01212}^{*}-\overline{\mathcal{A} }_{01221}^{*}-\overline{\tau }_{rr}\right), \quad  \overline{\mathcal{T} }=\left( \overline{\varGamma }_{0221}^{*}-\overline{\varGamma }_{0111}^{*} \right).
	\end{split}\tag{A2}
\end{align}

%
%
%

\bibliographystyle{unsrt}  
\bibliography{references}

\end{document}